\documentclass[11pt,a4paper]{article}
\usepackage{geometry}
\usepackage{amsthm}
\usepackage{amsmath}
\usepackage{ulem}
\usepackage{makecell}
\usepackage{tabularx}
\newcolumntype{P}[1]{>{\centering \arraybackslash}p{#1}}
\newcolumntype{L}{X}
\newcolumntype{C}{>{\centering \arraybackslash}X}
\newcolumntype{R}{>{\raggedright \arraybackslash}X}
\usepackage{graphicx}
\usepackage[nottoc]{tocbibind}
\usepackage{float}
\usepackage{bbold}
\usepackage{dutchcal}
\usepackage{graphicx}
\usepackage{tikz} 
\usepackage{amssymb}
\usepackage{graphics}
\usepackage{cite}
\usepackage{color, verbatim}
\usepackage{colortbl}

\newtheorem{example}{Example}

\def \cC {\mathcal{C}}

\def \cO {\mathcal{O}}

\def \1{{\mathbf{1}}}

\def \Max{\mathsf{Max}}
\def \p{\mathsf{p}}
\def \Min{\mathsf{Min}}
\def \Pr{\mathrm{Pr}}

\newcommand{\pp}[1] { \!\left( {#1} \right) }

\newcommand{\pn}[1] { \left| {#1} \right| }

\def\endproof {\rule{2mm}{2mm} \vspace*{3mm}}
\definecolor{darkgreen}{rgb}{0,0.7,0}
\definecolor{darkred}{rgb}{0.9,0,0}

\def \Old{\textcolor{blue}}

\begin{document}
\title{An improvement of Random Node Generator for the uniform generation of capacities}

\author{Peiqi SUN${}^{1}$\thanks{Corresponding author}, Michel GRABISCH${}^{2}$ and Christophe LABREUCHE${}^{3}$.\\
\normalsize $^{1}$ Universit\'e Paris I - Panth\'eon-Sorbonne, Paris, France\\
{\normalsize\tt peiqisun94@gmail.com}\\
\normalsize $^{2}$ Paris School of Economics, Universit\'e Paris I - Panth\'eon-Sorbonne, Paris, France  \\
{\normalsize\tt michel.grabisch@univ-paris1.fr}\\
\normalsize $^{3}$ Thales Research \& Technology, Palaiseau, France\\
{\normalsize\tt christophe.labreuche@thalesgroup.com}}

\date{\today}
\maketitle
\begin{abstract}
Capacity is an important tool in decision-making under risk and uncertainty and multi-criteria decision-making. When learning a capacity-based model, it is important to be able to generate uniformly a capacity. Due to the monotonicity constraints of a capacity, this task reveals to be very difficult. The classical Random Node Generator (RNG) algorithm is a fast-running speed capacity generator, however with poor performance. In this paper, we firstly present an exact algorithm for generating a $n$ elements' general capacity, usable when $n < 5$. Then, we present an improvement of the classical RNG by studying the distribution of the value of each element of a capacity. Furthermore, we divide it into two cases, the first one is the case without any conditions, and the second one is the case when some elements have been generated. Experimental results show that the performance of this improved algorithm is much better than the classical RNG while keeping a very reasonable computation time. 
\end{abstract}


\section{Introduction} 
Capacities and the Choquet integral are widely used in decision making, 
especially in decision with multiple criteria, where the capacity models the
importance of groups of criteria while the Choquet integral is used as a
versatile aggregation operator \cite{GrabischL,Grabisch16}. It is often useful
in practice to be able to randomly generate capacities, in a uniform way
(measure of performance of models, {evaluation of an elicitation technique},
learning/identification phase, etc.). This problem reveals to be surprisingly
difficult, because of the monotonicity constraints defining capacities, so that
naive approaches yield poor performance and give highly biased distributions.

The theoretical perfect solution to the random generation problem is however
known: since the set of capacities is an order polytope, generating capacities
in a uniform way amounts to generating all linear extensions of the Boolean
lattice $(2^N,\subseteq)$ \cite{Stanley}. {This leads to the exact capacity
  generator (ECG), which we have implemented.} However, the number of linear extensions of
$(2^N,\subseteq)$ grows tremendously fast with $n:=|N|$, and is even not known
beyond $n=8$. Therefore, approximate solutions have to be found. One way is to
generate a sufficiently representative subset of linear extensions: this is the
approach taken by Karzanov and Khachiyan using Markov Chains \cite{KK}, Combarro et
al \cite{Combarro13, Combarro19}, and
also the authors of this paper \cite{GCS}. Another way is to find some simple
heuristic for directly generating one by one all the coefficients of a capacity,
for example, the random node generator of Havens and Pinar \cite{HavensPinar}. This generator
is very fast but has poor performance, due to the fact that for simplicity the
coefficients of a capacity are supposed to follow a uniform distribution on some
interval. However, the theoretical distribution of a coefficient is very
complex and relies also on linear extensions.

The {first} aim of this paper is to provide an improvement of the random node
generator of Havens and Pinar, {called IRNG}, by taking advantage of some
properties of the exact distribution of the coefficients of a capacity. We show
that distributions obtained by our method are much closer to the exact
distributions or those obtained by the Markov Chain method, while demanding a
small computation time, which is much lower than the time required by the Markov
Chain method.

{Our second aim is motivated by the fact that, in practice, it is often
  necessary to generate a set of capacities in a uniform way but {\it subject to
    some constraints}, which could come from some preference information given
  by the decision maker, or when using the approach of Stochastic Multiobjective
  Acceptability Analysis (SMAA) \cite{lahoksal98,ancorgr13}. A naive solution to
  this problem is the acceptance and rejection method, which amounts to
  generating capacities and keeping only those which satisfy the
  constraints. Indeed, the acceptance rate would be in most cases too small to
  make the method tractable. Therefore, one has to take into account the
  constraints directly into the generation method.  Incorporating arbitrary
  linear constraints on capacity coefficients into ECG or IRNG seems however to
  be infeasible. We have therefore restricted to constraints where two capacity
  coefficients are compared, or one coefficient is compared to some value, and
  proposed a modification of ECG and IRNG in order to take into account these
  types of constraints. As preferential information leads in general to more
  complex constraints than the two types we restrict to, there is still a
  acceptance and rejection step in our approach, but with a much better
  acceptance rate than with the naive method.}

The paper is organized as follows: Section 2 gives the necessary background
  on multicriteria decision making, capacities, and the random generation of
  capacities. Section 3 presents the exact method based on linear extensions
  (ECG algorithm) and the improvement of the Random Node Generator (IRNG
  algorithm) based on the study of the theoretical distribution of the
  coefficients of a capacity. Section 4 is devoted to the experimental results
 on the comparison of IRNG with the original Random Node Generator and the
 Markov Chain method. In Section 5, we study how to incorporate constraints into
our ECG and IRNG algorithms. Experimental results are given to show the
advantage on the naive approach. Section 6 concludes the paper.

% ===================================================================================

\section{Background}
\label{Sback}

% -----------------------------------------------------------------------------------
\subsection{Multi-Criteria Decision Aiding}
\label{Sback1}

Multi-Criteria Decision Aiding (MCDA) consists in modeling the preferences of a Decision Maker (DM) regarding alternatives on the basis of multiple and conflicting criteria.
We denote by $N=\{1,\ldots,n\}$ the set of criteria.
Each {criterion} $i\in N$ is associated {with} an attribute $X_i$, and the alternatives are thus elements of the Cartesian product $X = X_1 \times \cdots \times X_n$.
The preference relation of the DM is denoted by $\succsim$ where $x \succsim y$ (for  $x,y\in X$) means that the DM prefers $x$ over $y$.
In general, we look for a {\it numerical representation} \cite{krlusutv71} $u:X\rightarrow
\mathbb{R}$ of the preference relation such that:
\begin{equation}
  \forall x,y\in X \ \ , \ \ x\succeq y \ \Leftrightarrow \ u(x) \geq u(y) .
\label{Eback1}
\end{equation}
Without loss of generality, we consider scale $[0,1]$, where $0$ (resp. $1$) means that the criterion is not satisfied at all (resp. perfectly satisfactory). Hence $u:X\rightarrow [0,1]$.
Function $u$ is often written in a decomposable {form} \cite{kera76}, where we first normalize the attributes by introducing a utility function mapping each attribute into the satisfaction scale $[0,1]$, and then aggregate the normalized scores to produce the overall score.
These two steps are in general handled separately, and for the purpose of this paper, we only consider the second step, that is the aggregation step.
Hence, in order to avoid cumbersome notation, we assume that the input scores are already normalized, that is
$X_1=\cdots=X_n=[0,1]$ and $X=[0,1]^N$.
Hence $u(x)=F(x_1,\ldots,x_n)$, where $F:[0,1]^n\rightarrow [0,1]$ is an {\it aggregation function}.

The most classical aggregation function is the weighted sum $F(x_1,\ldots,x_n) = \sum_{i\in N} w_i \: x_i$.
This model assumes that all criteria are independent, which is often violated in real applications.
We consider thus a more versatile aggregation function called the Choquet integral.
A {\it capacity} on $N$ is a set function $\mu:2^N \rightarrow [0,1]$ such that $\mu(\emptyset)=0$, $\mu(N)=1$ (normalization) and
$\mu(A) \leq \mu(B)$ whenever $A\subseteq B$ (monotonicity) \cite{Choquet}.
The set of capacities on $N$ is denoted by $\mathcal{C}(N)$.

The Choquet integral is a generalization of the weighted sum, taking into account the weights $\mu$ \cite{Choquet}: 
\begin{equation}
  C_\mu(x) = \sum_{k=1}^n \pp{ {x}_{\tau(k)} - {x}_{\tau(k-1)}} \: \mu(\{\tau(k),\ldots,\tau(n)\}) ,
\label{Eback2}
\end{equation}
where $\tau$ is a permutation on $N$ such that $x_{\tau(1)} \leq x_{\tau(2)} \leq \cdots \leq x_{\tau(n)}$, and with the notation $x_{\tau(0)} := 0$.
Capacity $\mu$ can be nicely interpreted as the overall score of a particular alternative: 
\begin{equation}
  C_\mu(1_B,0_{N\setminus B}) = \mu(B) ,
\end{equation}
where, for $x,y\in X$, $(x_B,y_{N\setminus B})$ denotes an alternative taking the value $x_i$ when $i\in B$ and $y_i$ otherwise.
Alternative $(1_B,0_{N\setminus B})$ is called {\it binary alternative} and is denoted by $a_B$.

Capacity $\mu$ is elicited from {\it preference information} provided by the DM.
It typically consists of
\begin{itemize}
\item a set $P$ of pairs of alternatives $(a,b) \in X^2$ such that $a$ is strictly preferred to $b$ for the DM, and 
\item a set $I$ of pairs of alternatives $(a,b) \in X^2$ such that $a$ is indifferent to $b$ for the DM.
\end{itemize}
We consider then the set of capacities that are compatible with the preference information $P,I$ of the DM:
\begin{align*}
 & \mathcal{C}(N;P,I) = \{ \mu\in \mathcal{C}(N) \ : \ 
   C_\mu(a) \geq C_\mu(b) + \varepsilon \ \forall (a,b) \in P
    \mbox{ and } C_\mu(a) = C_\mu(b) \ \forall (a,b) \in I \}
\end{align*}
where $\varepsilon>0$ is a fixed threshold.
{Usually, the capacity which is chosen in this set is} a solution of an
  optimization problem under constraint $\mathcal{C}(N;P,I)$ \cite{grkome06},
  where the objective function to maximize can be for example the entropy of $\mu$ \cite{komaro05}.

% -----------------------------------------------------------------------------------
\subsection{Need of random capacity generator}

The need to have a (unbiased) generator of capacities arise in many problems.
Genetic algorithms are classical techniques to learn a capacity \cite{Magoc2014}.
They start from a uniform population of capacities and require thus a random generator of capacities.
Another application is for the experimental evaluation of an elicitation technique \cite{herinUAI22}.
To this end, one needs a random generator of capacities, from which one can generate preference information.

Having a generator of (unbiased) capacities compatible with the preferential information is also quite important in decision.
This is {particularly} the case {with} the recommendation regarding a discrete set $A \subset X$ of options.
       Ideally, one would like to make a {\it robust} recommendation, which occurs when there exists $a\in A$ such that
        for all $b\in A\setminus \{a\}$ and all $\mu \in \mathcal{C}(N;P,I)$,  $C_\mu(a) \geq C_\mu(b)$ \cite{grmasl08,angcogrsl16}.
       Unfortunately, this condition is very strong and is far from being satisfied in most cases.
       A weaker version consists in counting the proportion of capacities in $\mathcal{C}(N;P,I)$ for which an alternative dominates another one, as depicted in Stochastic Multiobjective Acceptability Analysis (SMAA) \cite{lahoksal98,ancorgr13}.
       As this ratio cannot be computed theoretically, having a good numerical approximation of this quantity relies on an unbiased random generator of $\mathcal{C}(N;P,I)$. Note that SMAA traditionnally  uses the {\it Hit and Run} random generator of a polytope, which does not provide any guarantee to be unbiased.

% -----------------------------------------------------------------------------------
\subsection{Random generator of linear extensions}
\label{Sback3}

Let $P$ be a finite set, endowed with a partial order $\preccurlyeq$. We say that
$(P,\preccurlyeq)$ is a {\it (finite) poset}. We recall the following notions:
\begin{itemize}
\item $x\in P$ is {\it
maximal} if $x\preccurlyeq y$ with $y\in P$ implies $x=y$. We denote by
$\Max(P,\preccurlyeq)$ (simply $\Max(P)$) the set of maximal
elements of $P$. 
\item A {\it linear extension} of $(P,\preccurlyeq)$ is a total order $\leqslant$ on $P$ which
is compatible with the partial order $\preccurlyeq$ in the following sense:
$x\preccurlyeq y$ implies $x\leqslant y$. 
\item The {\it order polytope} \cite{Stanley}
associated to $(P,\preccurlyeq)$, denoted by $\cO(P)$, is the set
\[
\cO(P) = \{f:P\longrightarrow [0,1]\mid f(x)\leqslant f(y)\mbox{ if }
x\preccurlyeq y\}.
\] 
\end{itemize}
It is known from Stanley \cite{Stanley} that linear extensions induce a
triangulation of $\cO(P)$ into simplices of equal volume. Therefore, generating
in a random uniform way an element of $\cO(P)$ amounts to generating all linear
extensions, or to generating them randomly according to a uniform distribution.

We apply this result to capacities. It is easy to see that the set $\cC(N)$ of capacities is an order polytope, whose underlying poset is
$(2^N\setminus\{\varnothing,N\},\subseteq)$. Therefore, the problem of randomly
generating capacities according to a uniform distribution amounts to generating
the linear extensions of the poset $(2^N\setminus\{\varnothing,
N\},\subseteq)$. For example, for a 3 elements' capacity, $(\{1\}, \{2\}, \{3\},
\{1,2\}, \{1,3\}, \{2,3\})$ is a linear extension of the poset
$(2^{\{1,2,3\}}\setminus\{\varnothing,\{1,2,3\}\},\subseteq)$.
However, the number of linear extensions of $(2^N\setminus\{\varnothing,
N\},\subseteq)$ increases tremendously fast, and is unknown beyond $n=8$.

%====================================================================================
\section{Capacity Generators: Random Node generator based on Beta distribution}\label{sec:1}
\subsection{Exact Capacity Generator}

When $n\leq 4$, it is possible to have an exact algorithm generating all linear
extensions, and therefore to generate capacities in a uniform way. We propose
below such an algorithm ({\bf Exact-capacity-generator} (ECG)), which is recursive and performs a Depth-First-Search (DFS) finding maximal elements of a poset, which will
form the tail of the list describing the linear extension. The following
dendrogram of Figure 1 (right) illustrates the process of the algorithm for a 3
elements' capacity. The maximal element is $\{1,2,3\}$, which is the root of the dendrogram (Figure 1, right), then we continue to find the set of maximal elements of the poset deprived of node $\{1,2,3\}$, which is $\{\{1,2\},
\{1,3\}, \{2,3\}\}$, that is the second level of dendrogram. Next, we continue
to find the set of maximal elements when each node in the second level of the
dendrogram is removed. We repeat the above steps until there is only one element
left in the poset to obtain the whole dendrogram.

\begin{figure}[htb]
\begin{minipage}{0.5\linewidth}
\begin{tikzpicture}[scale = 1.7]
\node[right] at (-1,1){$1$};
\node[right] at (0,1){$2$};
\node[right] at (1,1){$3$};
\node[right] at (-1,2){$12$};
\node[right] at (0,2){$13$};
\node[right] at (1,2){$23$};
\node[right] at (0,0){$\emptyset$};
\node[right] at (0,3){$123$};
\draw[thick,->=0.5](0,0) -- (0,1); 
\draw[thick,->=0.5](0,0) -- (-1,1); 
\draw[thick,->=0.5](0,0) -- (1,1);
\draw[thick,->=0.5](0,1) -- (1,2);
\draw[thick,->=0.5](1,1) -- (1,2);
\draw[thick,->=0.5](-1,1) -- (-1,2);
\draw[thick,->=0.5](-1,1) -- (0,2);
\draw[thick,->=0.5](0,1) -- (-1,2);
\draw[thick,->=0.5](1,1) -- (0,2);
\draw[thick,->=0.5](-1,2) -- (0,3);
\draw[thick,->=0.5](0,2) -- (0,3);
\draw[thick,->=0.5](1,2) -- (0,3);
\draw[thick,->=0.5](0,0) -- (1,1);
\draw[thick,->=0.5](0,0) -- (-1,1);
\draw[thick,->=0.5](0,0) -- (1,1);
\draw[thick,->=0.5](0,0) -- (1,1);
\end{tikzpicture} 
\end{minipage}
\hfill
\begin{minipage}{.5\linewidth}
\begin{tikzpicture}[scale = 1.7]
\node[right] at (-1,2){$12$};
\node[right] at (0,2){$13$};
\node[right] at (1,2){$23$};
\node[right] at (0,3){$123$};
\draw[thick,=0.5](0,2) -- (0,3);
\draw[thick,=0.5](-1,2) -- (0,3);
\draw[thick,=0.5](1,2) -- (0,3);
\node[right] at (0,0.4){$\cdots$};
\node[right] at (1,0.4){$\cdots$};
\draw[thick,=0.5](-1.3,1) -- (-1,2);
\draw[thick,=0.5](-0.7,1) -- (-1,2);
\node[left] at (-1.3,1){$13$};
\node[left] at (-0.7,1){$23$};

\draw[thick,=0.5](-1.55,0) -- (-1.3,1);
\draw[thick,=0.5](-1.05,0) -- (-1.3,1);
\node[left] at (-1.55,0){$1$};
\node[left] at (-1.05,0){$23$};

\draw[thick,=0.5](-0.95,0) -- (-0.7,1);
\draw[thick,=0.5](-0.45,0) -- (-0.7,1);
\node[right] at (-0.95,0){$2$};
\node[right] at (-0.45,0){$13$};
\node[right] at (-1.5,-0.25){$\cdots$};
\node[right] at (-0.8,-0.25){$\cdots$};

\draw[thick,=0.5](-0.3,1) -- (0,2);
\draw[thick,=0.5](0.3,1) -- (0,2);
\node[right] at (-0.3,1){$12$};
\node[right] at (0.3,1){$23$};

\draw[thick,=0.5](0.7,1) -- (1,2);
\draw[thick,=0.5](1.3,1) -- (1,2);
\node[right] at (0.7,1){$12$};
\node[right] at (1.3,1){$13$};
\end{tikzpicture} 
\end{minipage}
\caption{Case $n=3$. Left: representation of the poset 
$(2^N,\subseteq)$. Right: Dendrogram of the maximal elements when running the procedure for generating all linear extensions using the DFS algorithm.} 
\label{zrotate}
\end{figure}
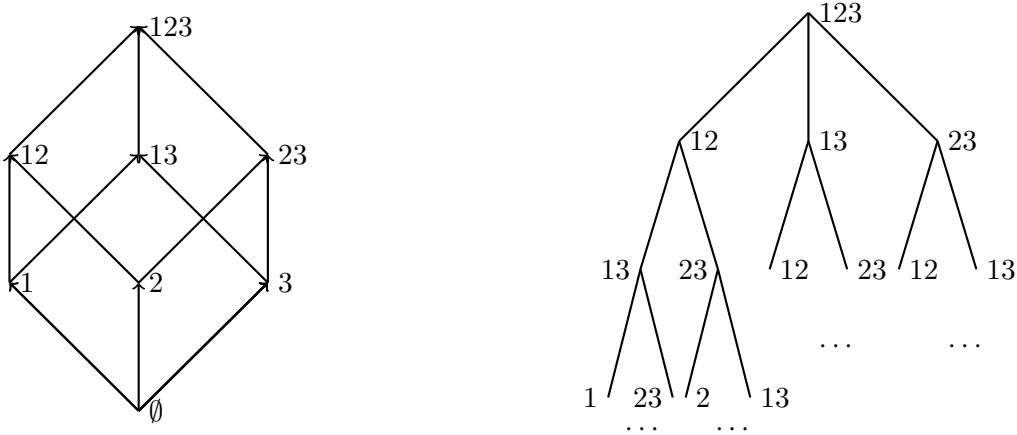

\begin{quote}
\hrulefill\ \raisebox{-.2\baselineskip}{Algorithm 1}\hrulefill \par
{\bf Exact-capacity-generator}$(n,k)$ \par
\hrulefill \par 
{\bf Input:} $n,k$ integers.\\
$\%$ $k$ is the number of all linear extensions and $n = |N|$.\\
{\bf Output:} $k$ generated capacities on $2^N$\\
\% $AllLinear$ is an empty array which will contain all linear extensions\\
1: $count \leftarrow 0$\\
{\hspace*{0.3cm}}\% $P$ is an array containing the poset $2^N\setminus\{\varnothing,N\}$\\
2: {\bf All-linear-extension}($P,AllLinear,count$) \\
3: {\bf repeat $k$ times}\\
4: {\hspace*{0.7cm}}{Select uniformly one linear extension of $AllLinear$}\\
5: {\hspace*{0.7cm}} Generate uniformly $2^n-2$ numbers between 0 and 1, sort them from smallest to  {\hspace*{0.95cm}} largest,  and assign them to the selected linear extension\\
{\bf end repeat} \par
\hrulefill
~\\

\hrulefill\ \raisebox{-.2\baselineskip}{Algorithm 2}\hrulefill \par
{\bf All-linear-extension}$(P,AllLinearExtensions,count)$ \par
\hrulefill \par 
$\%$$AllLinearExtensions$ stores all linear extensions of $P$ and $count$ stores the number of linear extensions \\
{\bf Input:} an array $P$ containing a poset of size $n$, an array $AllLinearExtensions$ and $count$ \\
{\bf Output:} All linear extensions of poset $P$\\
6: {\bf If} $|P| = 1$ {\bf then} \\
{\hspace*{1cm}} $\%$ When the bottom of dendrogram is reached, add an empty linear extension to $AllLinearExtensions$. \\
7: {\hspace*{0.7cm}} Append a zeros array of size $n$ to $AllLinearExtensions$ \\
8: {\hspace*{0.7cm}} $AllLinearExtensions[count-1][n-1] \leftarrow P[0]$ \\
9: {\hspace*{0.7cm}} $count \leftarrow count +1$ \\
{\hspace*{0.3cm}}{\bf end if}\\
10: {\bf For} i in $\Max(P)$ {\bf do}\\
11: {\hspace*{0.7cm}} Remove i from $P$\\
{\hspace*{1.2cm}} $\%$ recursion algorithm \\
12: {\hspace*{0.7cm}} {\bf All-linear-extension}($P,AllLinearExtensions,count$)\\
13: {\hspace*{0.7cm}} $AllLinearExtensions[count-1][\text{size of }P] \leftarrow$ i \\
14: {\hspace*{0.7cm}} Re-insert i to the end of poset $P$ \\
{\hspace*{0.3cm}}{\bf end for}

\hrulefill
\end{quote}

When $n>4$, approximate methods have to be used, either generating randomly
linear extensions like the Markov Chain method \cite{KK}, the 2-level
approximation method \cite{GCS}, etc., or based on other principles
like the {\it Random Node generator (RNG) algorithm} introduced by T. C. Havens
and A. J. Pinar in \cite{HavensPinar}. The core idea of this approach is to
randomly select one element $S\in 2^N$ among all elements and then draw it with
a uniform law between the maximum and minimum values allowed by the
monotonicity constraints. This operation is repeated until all elements in $2^N$
have assigned values.

The most significant advantage of this method is its low complexity and fast
running speed. However, theoretically, the capacities generated by it are not
uniform, because firstly the range of values for $\mu(S)$ is highly dependent on
the rank in which the element $S$ is selected, and secondly the exact
distribution of $\mu(S)$ is far from being a uniform distribution. Therefore,
this capacity generator has a lot of theoretical undesirabilities.

As an illustration, we compare the performance of the RNG with ECG. The following figures show the distribution of $\mu(S)$ generated by the RNG and ECG.

\begin{figure}[htb] 
\centering 
\includegraphics[width=1\textwidth]{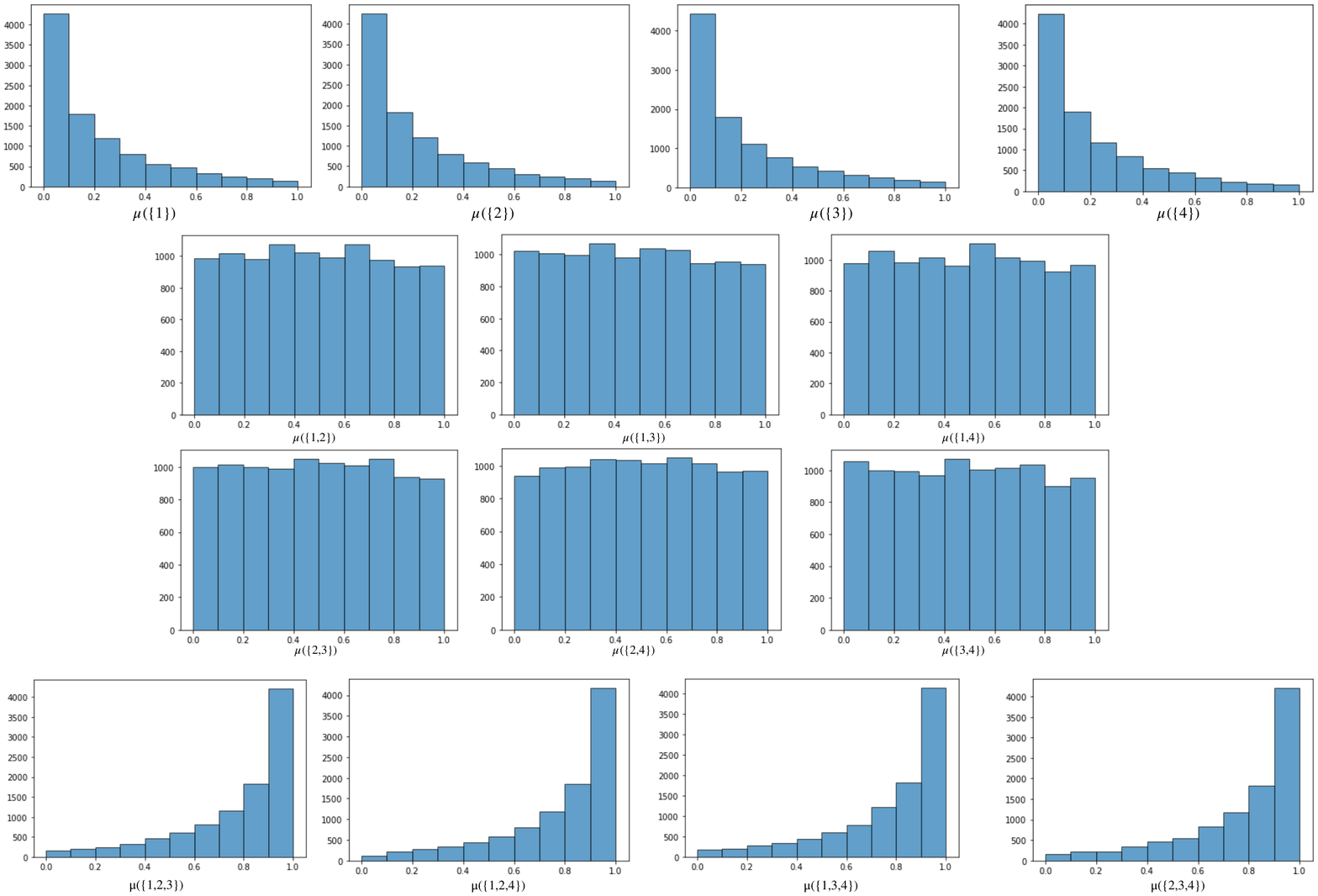}
\caption{Case $n=4$. Histograms of the values of $\mu(S)$,
$S\in 2^N\setminus\{N,\varnothing\}$, 
generated by RNG (compare with Fig.~\ref{fig:eg} where the exact generator
has been used).}
\label{fig:rn}
\end{figure}
\begin{figure}[htbp] 
\centering 
\includegraphics[width=1\textwidth]{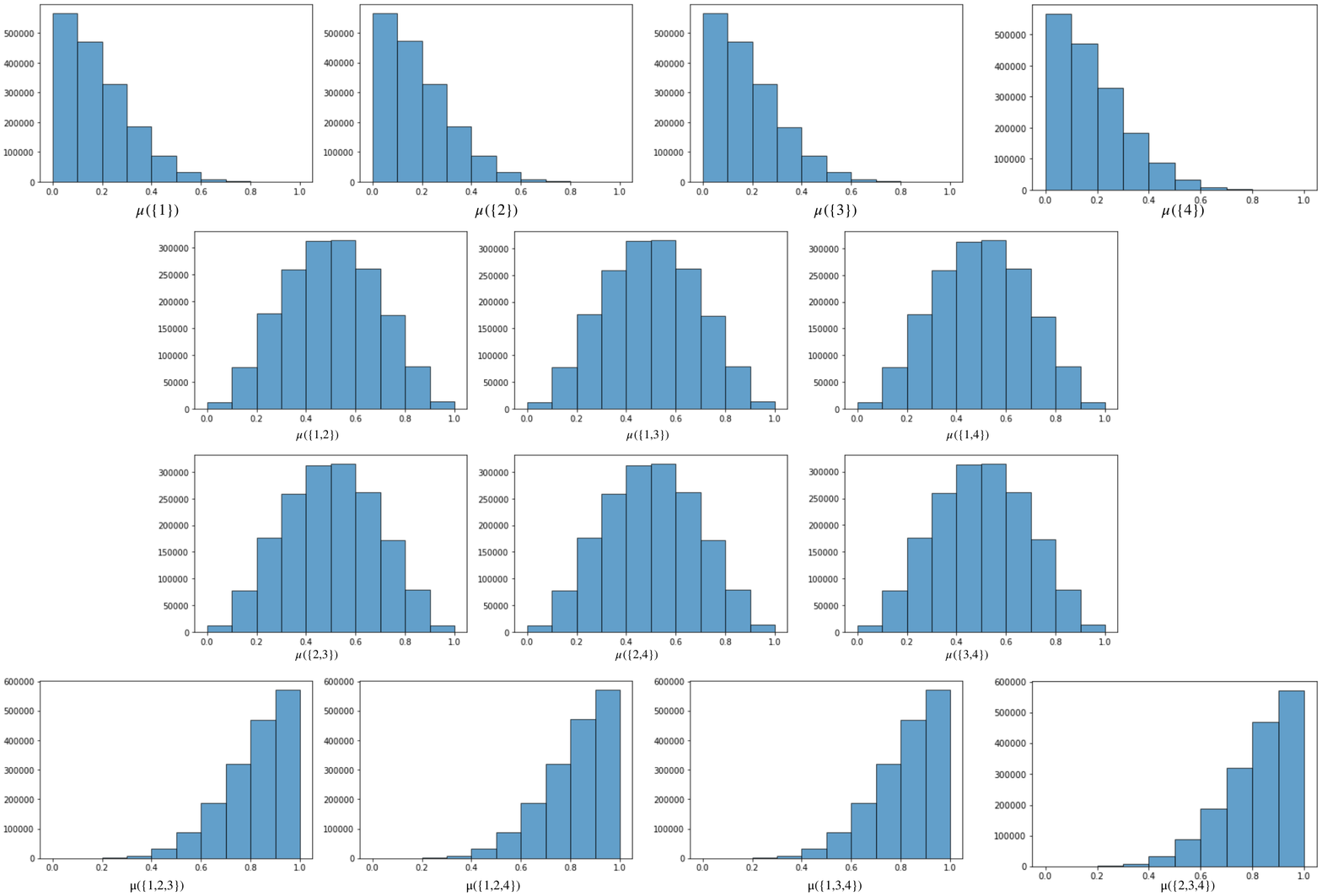} 
\caption{Case $n=4$. Histograms of the values of $\mu(S)$,
$S\in 2^N\setminus\{N,\varnothing\}$, 
generated by ECG. }
\label{fig:eg}
\end{figure}

From Figures~\ref{fig:rn} and \ref{fig:eg}, we notice that the discrepancy
between the distribution of these two groups of $\mu$ is significant, and thus
we may conclude that the uniformity of the capacity obtained by the Random-Node
generator is not satisfactory. In the next subsections, we study the theoretical
distribution of $\mu(S)$ and propose an improvement of the Random-Node
generator.

\subsection{Theoretical distribution of $\mu$}
\label{sec:thdi}
{ The main idea for improving the random node generator algorithm is to use a more realistic probabilistic distribution on the generation of the capacity of the current subset $S$.
Let us first describe the probability distribution of such a term.

To this end, let us consider } a set of i.i.d random variables $\mu_1, \mu_2, \ldots, \mu_m$ {\ that follow} the uniform law between 0 and 1.
We sort the $\mu_{i}$s into the order statistics $\mu_{(1)} \leq \mu_{(2)} \cdots \leq \mu_{(m)}$.
Then $\mu_{(k)}$ follows the Beta distribution $\mu_{(k)} \sim \operatorname{Beta}(k,m-k+1)$. If we take $\alpha = k, \beta = m-k+1$, then the formula for the density of $\mu_{(k)}$ is as follows:
\[ f_{\mu_{(k)}}(x)= \frac{\Gamma(\alpha+\beta)}{\Gamma(\alpha)\Gamma(\beta)}x^{\alpha-1}(1-x)^{\beta-1} ,
\]
where $\Gamma(\alpha) = \int_0^{\infty}t^{\alpha-1}e^{-t}dt,$ with $\alpha >0$.

{\ In order to apply this result to capacities, we need to know the rank of $\mu(S)$ within all terms of a capacity.}
We denote by $\mathcal{Rk}(S)$ the rank of element $S$ ($S \in 2^N$) in a linear extension of poset $(2^N,\subseteq)$. 
Each element of $2^N$ has a rank in each linear extension corresponding to the poset, among them $\varnothing$ is always located at the minimal rank, i.e., $\mathcal{Rk}(\varnothing) = 0$ and $N$ is always located at the maximal rank, i.e., $\mathcal{Rk}(N) = 2^n-1$. 

{Then the cumulative distribution function of $\mu(S)$, i.e. $ \mathbb{P}(\mu(S)\leq x)$ for $0<x<1$, considers the beta distribution over all possible rankings of $\mu(S)$}
\begin{align*}
& F_{\mu(S)}(x) = \mathbb{P}(\mu(S)\leq x) = \sum_{i = \min(\mathcal{Rk}(S))}^{\max(\mathcal{Rk}(S))} \mathbb{P}(\mu(S)\leq x | \mathcal{Rk}(S) = i)\times \mathbb{P}(\mathcal{Rk}(S) = i) \\
& = \sum_{i = \min(\mathcal{Rk}(S))}^{\max(\mathcal{Rk}(S))} \mathbb{P}(\mu_{(i)}\leq x) \times \mathbb{P}(\mathcal{Rk}(S) = i) \\
& = \sum_{i = \min(\mathcal{Rk}(S))}^{\max(\mathcal{Rk}(S))} F_{\mu_{(i)}}(x)\times \mathbb{P}(\mathcal{Rk}(S) = i) ,
\end{align*}
with $F_{\mu_{(i)}}(x)$ the cumulative distribution function of Beta$(i, 2^n-1-i)$,
{\ $\min(\mathcal{Rk}(S))$ the smallest possible ranking of $\mu(S)$
and $ \max(\mathcal{Rk}(S))$ the largest possible ranking of $\mu(S)$.
These bounds on the ranking of $\mu(S)$ are simply obtained by the monotonicity condition, counting the minimal number of terms ranked before and after $\mu(S)$.
We obtain
\begin{align*}
& \min(\mathcal{Rk}(S)) = |\{ T \subseteq S \: , T\not= \emptyset\}| = 2^{|S|}-1 \\
& \max(\mathcal{Rk}(S)) = 2^n - |\{ T \supseteq S \ , \: T \subseteq N\}| = 2^n - 1-2^{|N\setminus S|} .
\end{align*}
}
The density of $\mu(S)$ is thus:
\begin{equation}
f_{\mu(S)}(x) = \sum_{i = min(\mathcal{Rk}(S))}^{max(\mathcal{Rk}(S))} f_{\mu_{(i)}}(x) \times \mathbb{P}(\mathcal{Rk}(S) = i) 
\label{EqDensMuS}
\end{equation} 
with $\mu_{(i)} \sim \operatorname{Beta}(i, 2^n-1-i)$. 

\medskip

{Density (\ref{EqDensMuS}) is correct when $\mu(S)$ is not constrained by other terms of the capacity.}
When we use the RNG to generate a capacity, we should adjust the above distribution due to its monotonicity. Supposing we have already generated the elements $S_1,\ldots,S_p$ with the values $\mu(S_1)=a_1,\ldots,\mu(S_p)=a_p$, 
{\ we wish to draw the distribution of $\mu(S)$ for a new subset $S$.
Compared to (\ref{EqDensMuS}), the knowledge of $a_1,\ldots,a_p$ provides constraints on both the numerical value of $\mu(S)$ and also its ranking.
Following the monotonicity conditions, we first note that the value of $\mu(S)$ shall belong to interval $[ \Min_{\p} \mu(S) , \Max_{\p} \mu(S)]$ where
\[ \Min_{\p} \mu(S) = \max_{ j\in \{1,\ldots,p\} \, , \: S_j \subseteq S} a_j \ \mbox{ and } \ \Max_{\p} \mu(S) = \min_{j\in \{1,\ldots,p\} \, , \: S_j \supseteq S} a_j .
\]
Moreover, as illustrated by the following example, the smallest and largest possible rankings of $\mu(S)$ are also constrained by $a_1,\ldots,a_p$.

\begin{example}
Assume that we have already generated the following terms $\mu(\{1,2\})=0.1$, $\mu(\{1,3\})=0.2$ and $\mu(\{4,5\})=0.3$, and consider now $S=\{1,4,5\}$ with $N=\{1,2,3,4,5\}$.
Subset $\{1,2\}$ and all its subsets are thus ranked before $\{4,5\}$. The same holds for $\{1,3\}$. 
In total, the subsets that are necessarily ranked before $S$ are the following: $\{1\}, \{2\}, \{1,2\}, \{3\}, \{1,3\}, \{4\}, \{5\}, \{4,5\}, \{1,4\}, \{1,5\}$.
Hence $S$ has rank at least $11$.
\endproof
\label{Ex1}
\end{example}

Generalizing the previous example, 
\[ \underline{\mathcal{S}}_p(S)=\{ S_j \ , \: j\in \{1,\ldots,p\} \ \mbox{s.t. } \exists i \in \{1,\ldots,p\} \: , \ S_i \subseteq S \mbox{ and } a_j \leq a_i \} \cup \{S\}
\]
is the set of already generated subsets that are necessarily ranked before $S$ (including $S$), and
\[ \overline{\mathcal{S}}_p(S)=\{ S_j \ , \: j\in \{1,\ldots,p\} \ \mbox{s.t. } \exists i \in \{1,\ldots,p\} \: , \ S_i \supseteq S \mbox{ and } a_j \geq a_i \} \cup \{S\}
\]
is the set of already generated subsets that are necessarily ranked after $S$ (including $S$).
The smallest possible ranking $\Min_{\p}\mathcal{Rk}(S)$ of $S$ is thus given by the number of subsets of $\underline{\mathcal{S}}_p(S)$.
It is not simply the sum of the subsets of the elements of $\underline{\mathcal{S}}_p(S)$ as there are common subsets.
In Ex. \ref{Ex1}, subset $\{1\}$ is a subset of $\{1,2\}$, $\{1,3\}$ and
$\{1,4,5\}$, and it shall not be counted three times.
To this end, we use the Poincar\'e sieve formula.
This formula provides the number of elements of the union of an arbitrary number of sets:
\[ \pn{\cup_{i=1}^q A_i } = \sum_{k=1}^q \pp{ (-1)^{k-1} \sum_{1\leq i_1 < i_2 < \cdots < i_k \leq q} \pn{ A_{i_1}\cap A_{i_2}\cap \cdots \cap A_{i_k}} } .
\]
We apply this formula to 
$A_j = 2^{\underline{S}_j}\setminus \{\emptyset\}$,
where $\underline{\mathcal{S}}_p(S) := \{ \underline{S}_1,\ldots,\underline{S}_{q} \}$.
As $A_{i_1}\cap \cdots \cap A_{i_k} = 2^{\underline{S}_{i_1} \cap \cdots \cap \underline{S}_{i_k}}\setminus\{\emptyset\}$, we obtain
\begin{align}
& \Min_{\p}\mathcal{Rk}(S) = \pn{ \{ T \subseteq \underline{S}_j, \ T\not=\emptyset \mbox{ and } j\in \{1,\ldots,q\} } \nonumber \\
& = \sum_{k=1}^q \pp{ (-1)^{k-1} \sum_{1\leq i_1 < i_2 < \cdots < i_k \leq q} \pp{2^{\pn{ \underline{S}_{i_1}\cap \underline{S}_{i_2}\cap \cdots \cap \underline{S}_{i_k}}} -1} } . \label{EqMinRk2}
\end{align}

\begin{example}[Ex. \ref{Ex1} continued]
We obtain $\Min_{\p} \mu(\{1,4,5\}) = 0.3$.
Moreover, we have $\underline{\mathcal{S}}_p(\{1,4,5\}) = \big\{ \{1,2\}, \{1,3\}, \{4,5\}, \{1,4,5\} \big\}$.
Applying (\ref{EqMinRk2}), the smallest possible ranking of $\{1,4,5\}$ is
\begin{align*}
& \pp{2^{|\{1,2\}|}-1} + \pp{2^{|\{1,3\}|}-1} + \pp{2^{|\{4,5\}|}-1} + \pp{2^{|\{1,4,5\}|}-1} \\
& - \pp{2^{|\{1\}|}-1} - \pp{2^{|\{1\}|}-1} - \pp{2^{|\{1\}|}-1} - \pp{2^{|\{4,5\}|}-1} + \pp{2^{|\{1\}|}-1} \\
& = 3+3+3+7-1-1-1-3+1 =11 . 
\end{align*}
Hence we recover that $S$ has rank at least $11$.
\endproof
\label{Ex2}
\end{example}

Likewise, the largest possible ranking $\Max_{\p}\mathcal{Rk}(S)$ of $S$ is given by
\[ 2^n - 1-\pn{ \{ T \supseteq \overline{S}_j, \ T\not=N \mbox{ and } j\in \{1,\ldots,q'\} } ,
\]
where $\overline{\mathcal{S}}_p(S) := \{ \overline{S}_1,\ldots,\overline{S}_{q'} \}$.
Applying the Poincar\'e sieve formula to $A_j = \{ T \subseteq \overline{S}_j \: , \ T \not= N\}$, we obtain
$|A_{i_1}\cap \cdots \cap A_{i_k}| = |\{ T \supseteq \overline{S}_{i_1},\ldots,\overline{S}_{i_k} \: , \ T \not= N\}| = |\{ T \supseteq \overline{S}_{i_1} \cup \cdots \cup \overline{S}_{i_k} \: , \ T \not= N\}|
= 2^{|N \setminus ( \overline{S}_{i_1} \cup \cdots \cup \overline{S}_{i_k})|}-1$ and 
\begin{align}
& \Max_{\p}\mathcal{Rk}(S) = 2^n - 1-\pn{ \{ T \supseteq \overline{S}_j, \ T\not=N \mbox{ and } j\in \{1,\ldots,q'\} } \nonumber \\
& = 2^n - 1- \sum_{k=1}^{q'} \pp{ (-1)^{k-1} \sum_{1\leq i_1 < i_2 < \cdots < i_k \leq q'} \pp{2^{|N \setminus ( \overline{S}_{i_1} \cup \cdots \cup \overline{S}_{i_k})|}-1} } . \label{EqMinRk3}
\end{align}

\begin{example}
Assume that we have already generated the following terms $\mu(\{1,2,3\})=0.9$, $\mu(\{1,3,4\})=0.8$ and $\mu(\{1,2,4,5\})=0.7$, and consider now $S=\{1,2,5\}$ with $N=\{1,2,3,4,5\}$.
We obtain $\Max_{\p} \mu(\{1,2,5\}) = 0.7$.
Moreover, we have $\overline{\mathcal{S}}_p(\{1,2,5\}) = \big\{ \{1,2,3\}, \{1,3,4\}, \{1,2,4,5\}, \{1,2,5\} \big\}$.
The subsets (excluding $N$) ranked after $\{1,2,5\}$ are
$\{1,2,5\}, \{1,2,3,5\}, \{1,2,4,5\}, \{1,2,3\}, \{1,2,3,4\}, \{1,3,4\}, $ 
$\{1,3,4,5\}$.
We obtain $7$ subsets.

Applying (\ref{EqMinRk3}), the largest possible ranking of $\{1,2,5\}$ is
\begin{align*}
& 2^5 - 1- \pp{2^{|\{3,4\}|}-1} - \pp{2^{|\{4,5\}|}-1} - \pp{2^{|\{2,5\}|}-1} - \pp{2^{|\{3\}|}-1} \\
& + \pp{2^{|\{5\}|}-1} + \pp{2^{|\{4\}|}-1} - \pp{2^{|\{3\}|}-1} \\
& = 2^5 - 1-3 - 3 - 3 - 1 + 3 = 2^n -1-7 = 24 
\end{align*}
\endproof
\label{Ex3}
\end{example}
}

Summarizing, the distribution of $\mu(S)$ becomes a conditional distribution:

\begin{align}
& \mathbb{P}( \mu(S) \leq x | \mu(S_1)=a_1,\ldots,\mu(S_p)=a_p ) \label{Eqq1} \\
& = \sum_{i = min_p\mathcal{Rk}(S)}^{max_p\mathcal{Rk}(S)} \mathbb{P}( \mathcal{Rk}(S)=i| \mu(S_1)=a_1,\ldots,\mu(S_p) = a_p) \nonumber \\
& \qquad \qquad \qquad \times \mathbb{P}( \mu(S) = \mu_{(i)} \leq x | \mu(S_1)=a_1,\ldots,\mu(S_p)=a_p)
\nonumber
\end{align}
with
\begin{align}
& \mathbb{P}(\mathcal{Rk}(S)=i | \mu(S_1)=a_1,\ldots,\mu(S_p)=a_p) \nonumber \\
& \approx \mathbb{P}( \mathcal{Rk}(S)=i | \Min_{\p}\mathcal{Rk}(S) \leq
\mathcal{Rk}(S) \leq \Max_{\p}\mathcal{Rk}(S) )\label{Eqq2}
\end{align}
and
\begin{align}
& \mathbb{P}( \mu(S) = \mu_{(i)} \leq x | \mu(S_1)=a_1,\ldots,\mu(S_p)=a_p ) \nonumber \\
& = \mathbb{P}( \mu_{(i)} \leq x | \Min_{\p} \mu(S) \leq \mu_{(i)} \leq
\Max_{\p} \mu(S))\label{Eqq3}
\end{align}

%==============================================================
\subsection{The improved random node generator}
Thanks to the previous considerations and Equations~(\ref{Eqq1}), (\ref{Eqq2})
and (\ref{Eqq3}), we are in a position to propose an improvement of the random
node generator, which we call IRNG. 

As explained, our improvement consists in replacing the uniform distribution of
$\mu(S)$ in the interval $[\Min_{\p} \mu(S),\Max_{\p} \mu(S)]$ by the
distribution given by (\ref{Eqq1}), computed through (\ref{Eqq2}) and
(\ref{Eqq3}). 

According to Equation (\ref{Eqq3}), when we assign a value to $\mu(S)$, it
should be between $\Min_{\p} \mu(S)$ and $\Max_{\p}\mu(S)$. If this is not
satisfied, we need to reject it and reassign a new value to $\mu(S)$. 

As for Equation~(\ref{Eqq2}), it is necessary to know the probability {$\mathbb{P}(\mathcal{Rk}(S) = i)$ for a
given subset $S$ to be ranked at $i$th position in a linear extension. This probability is stored in array $probability$ (where $probability[S][i]=\mathbb{P}(\mathcal{Rk}(S) = i)$) in the following algorithm.}
As the set of linear extensions is not practically reachable beyond $n=5$ and not known
beyond $n=8$, no practical expression of this probability can be obtained, and
it must be estimated. Therefore, the critical issue for the precision of the IRNG
algorithm is how to get these probabilities. Our proposition is to use off line
some well-performing method to generate randomly in a uniform way linear
extensions of $(2^N\setminus\{\varnothing, N\},\subseteq)$, like the Markov
chain method \cite{KK}, generating a sufficient number of linear
extensions from which $\mathbb{P}(\mathcal{Rk}(S) = i)$ could be estimated, for
every subset $S$ and every rank $i$. Once we have obtained these probabilities, we store them in a file so that they can be used repeatedly.

\begin{quote}
\hrulefill\ \raisebox{-.2\baselineskip}{Algorithm 3}\hrulefill \par
{\bf Improved-Random-Node-generator (IRNG)}$(P,probability)$ \par
\hrulefill \par 
{\bf Input:} a poset $P$ of $2^N\setminus\{\varnothing,N\}$, a two dimensional
array named $probability[S][j]$ containing the probability of element (subset)
$S\in 2^N\setminus\{\varnothing,N\}$ to be
at rank $j$. \\
{\bf Output:} capacity $\mu$ in $\mathcal{C}(N)$ generated with approximation method \\ 
1: AssignedElement,AssignedValue $\leftarrow$[ ],[ ]\\
2: $\mu \leftarrow$ a zero array of size $2^n-2$ \\
{\hspace*{0.3cm}}$\%$AssignedElement and AssignedValue store the elements
$S_1,\ldots, S_p$ \\
{\hspace*{0.3cm}}$\%$and element's value $a_1,\ldots, _p$ that have been already assigned \\
3: $\mathcal{L} \leftarrow$ an array of elements of $2^N\setminus\{\varnothing,N\}$ in random order\\
4: $p \leftarrow 0$\\
5: {\bf for} $S$ in $\mathcal{L}$ {\bf do}\\ 
6: {\hspace*{0.7cm}}Compute $\Min_{\p} \mu([S]), \Max_{\p} \mu([S])$ and $\Min_{\p}\mathcal{Rk}(S)$, $\Max_{\p}\mathcal{Rk}(S)$\\
{\hspace*{1cm}}$\% \ \Min_{\p}\mathcal{Rk}(S)$, $\Max_{\p}\mathcal{Rk}(S)$ the ranking restrictions of $S$ \\ 
{\hspace*{1cm}}\% and $\Min_{\p} \mu([S]), \Max_{\p} \mu([S])$ the minimum and maximum value of $\mu([S])$\\
7: {\hspace*{0.7cm}}$beta \leftarrow 0$\\
8: {\hspace*{0.7cm}}$\Pr_{min} \leftarrow \sum_{j=0}^{\Min_{\p}\mathcal{Rk}(\mu([S]))-1} probability[S][j]$\\
9: {\hspace*{0.7cm}}$\Pr_{max} \leftarrow \sum_{j=\Max_{\p}\mathcal{Rk}(\mu([S]))+1}^{2^n-3} probability[S][j]$\\
10:{\hspace*{0.6cm}}{\bf While} $beta \geq \Max_{\p}\mu([S])$ {\bf or} $beta \leq \Min_{\p}\mu([S])$ {\bf do}\\
{\hspace*{1cm}}$\%$ Capacity should obey monotonicity \\
11: {\hspace*{1.7cm}}$r \sim U([0,1])$\\
12: {\hspace*{1.7cm}}$r \leftarrow \Pr_{min}+(1-\Pr_{max}-\Pr_{min})*r$\\
13: {\hspace*{1.7cm}}$Rank \leftarrow \Min_{\p}\mathcal{Rk}(S))$\\
14: {\hspace*{1.7cm}}$\Pr \leftarrow \Pr_{min}$\\
15: {\hspace*{1.7cm}}{\bf While} $r > \Pr$ {\bf do}\\
16: {\hspace*{2.7cm}}$\Pr \leftarrow \Pr + probability[S][Rank]$\\
17: {\hspace*{2.7cm}}$Rank \leftarrow Rank+1$ \\
{\hspace*{2.2cm}}{\bf end while}\\
18: {\hspace*{1.7cm}}$beta \sim \operatorname{Beta}(Rank, 2^n-1-Rank)$. \\
{\hspace*{1.2cm}}{\bf end while}\\
19: {\hspace*{0.6cm}}$\mu[S] \leftarrow beta$ \\
20: {\hspace*{0.6cm}}Append $\mu[S]$ to Assignedvalue \\
21: {\hspace*{0.6cm}}Append $S$ to AssignedElement \\
22: {\hspace*{0.6cm}}$p \leftarrow p+1$\\
{\hspace*{0.1cm}} {\bf end for}

\hrulefill
\end{quote}

Let us analyze the computational complexity of one run to IRNG.
The $2^n-2$ subsets are ordered in array $\mathcal{L}$.
In $l.5$, we sweep these elements with an index $p$ from $p=1$ to $p=2^n-2$. At iteration $p$ ($l.6-22$), the complexity is given by the successive steps:
\begin{itemize}
\item $l.6$: the computation of $\Min_{\p} \mu([S])$ and $\Max_{\p} \mu([S])$ requires $p$ operations;
\item $l.6$: the computation of $\Min_{\p}\mathcal{Rk}(S)$ and $\Max_{\p}\mathcal{Rk}(S)$ requires $2^q+2^{q'} \leq 2^p$ operations (see (\ref{EqMinRk2}) and (\ref{EqMinRk3}));
\item $l.8-9$: the computation of $\Pr_{min}$ and $\Pr_{max}$ requires at most $2^n$ operations;
\item We assume that the While loop in $l.10$ is run at most $M$ times. The While loop in $l.15$ is run at most $2^n$ times.
Then the complexity of $l.10-18$ is $M \times 2^n$.
\end{itemize}
In total, the complexity of one run of IRNG is $O(2^n)$.
The main uncertainty in the computation time is the number of times $M$ the While loop in $l.10$ is run.
In the worst case, it could be large if interval $[\Min_{\p}\mu([S]), \Max_{\p}\mu([S])]$ is very small and $Rank$ is not well adapted to this interval.
This situation occurs with a low probability.

\section{Experimental results for IRNG}

\begin{figure}[htb] 
\centering 
\includegraphics[width=1\textwidth]{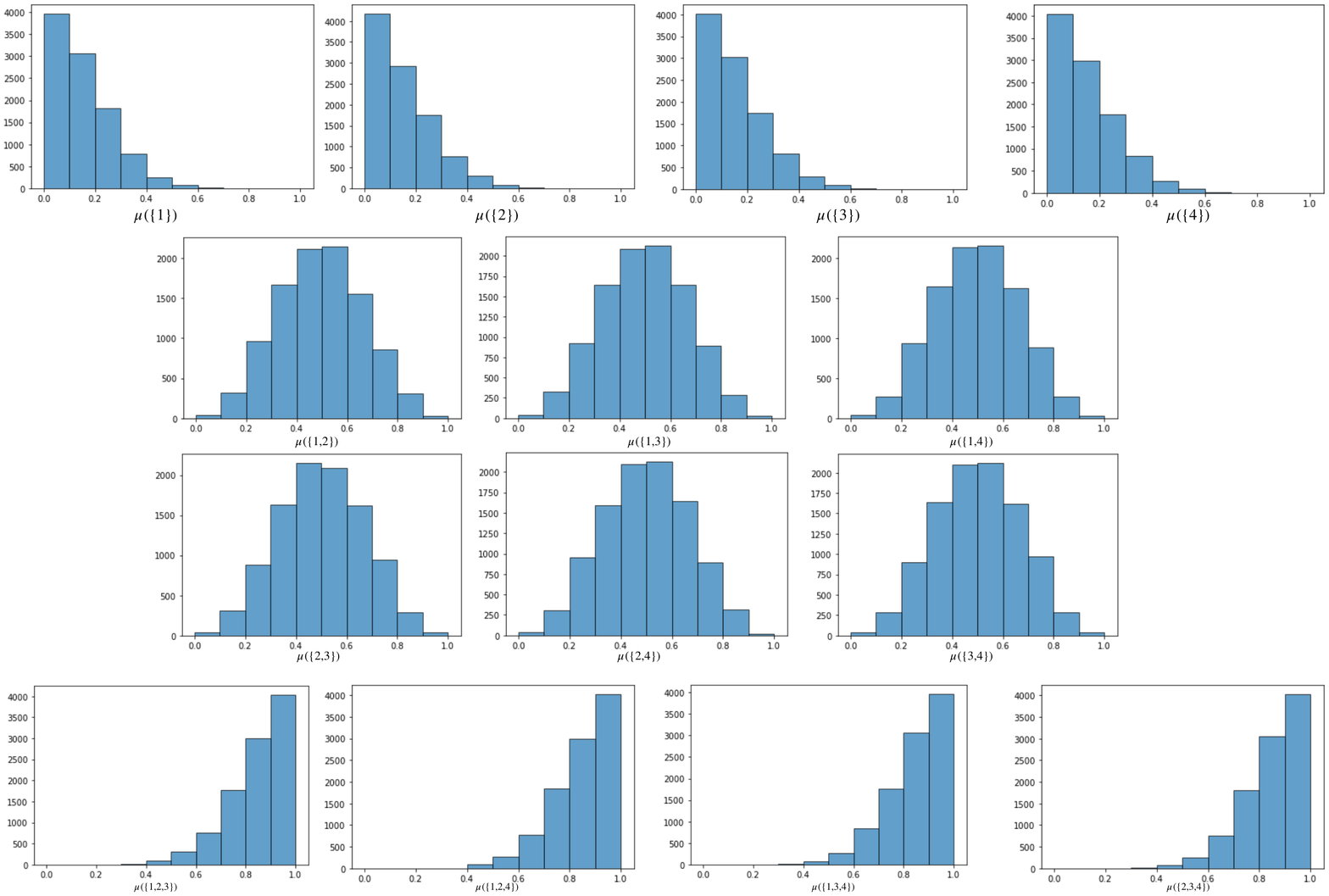}
\caption{Case $n=4$. Histograms of the values of $\mu(S)$,
$S\in 2^N\setminus\{N,\varnothing\}$, 
generated by IRNG (compare with Fig.~\ref{fig:eg} where the exact generator
has been used).}
\label{fig:irn}
\end{figure}

We compare the performance of the IRNG with the
RNG and Markov Chain generator. We apply the Markov
Chain method to obtain $\mathbb{P}(\mathcal{Rk}(S) = i)$ for all the following
experiments. In the experiments, we limit ourselves to $n=4$ in order to be able to compare the results with the ECG.
Figure~\ref{fig:irn} shows the distribution of $\mu(S)$ generated
by the IRNG for $n=4$.

\begin{figure}[htb] 
\centering 
\includegraphics[width=1\textwidth]{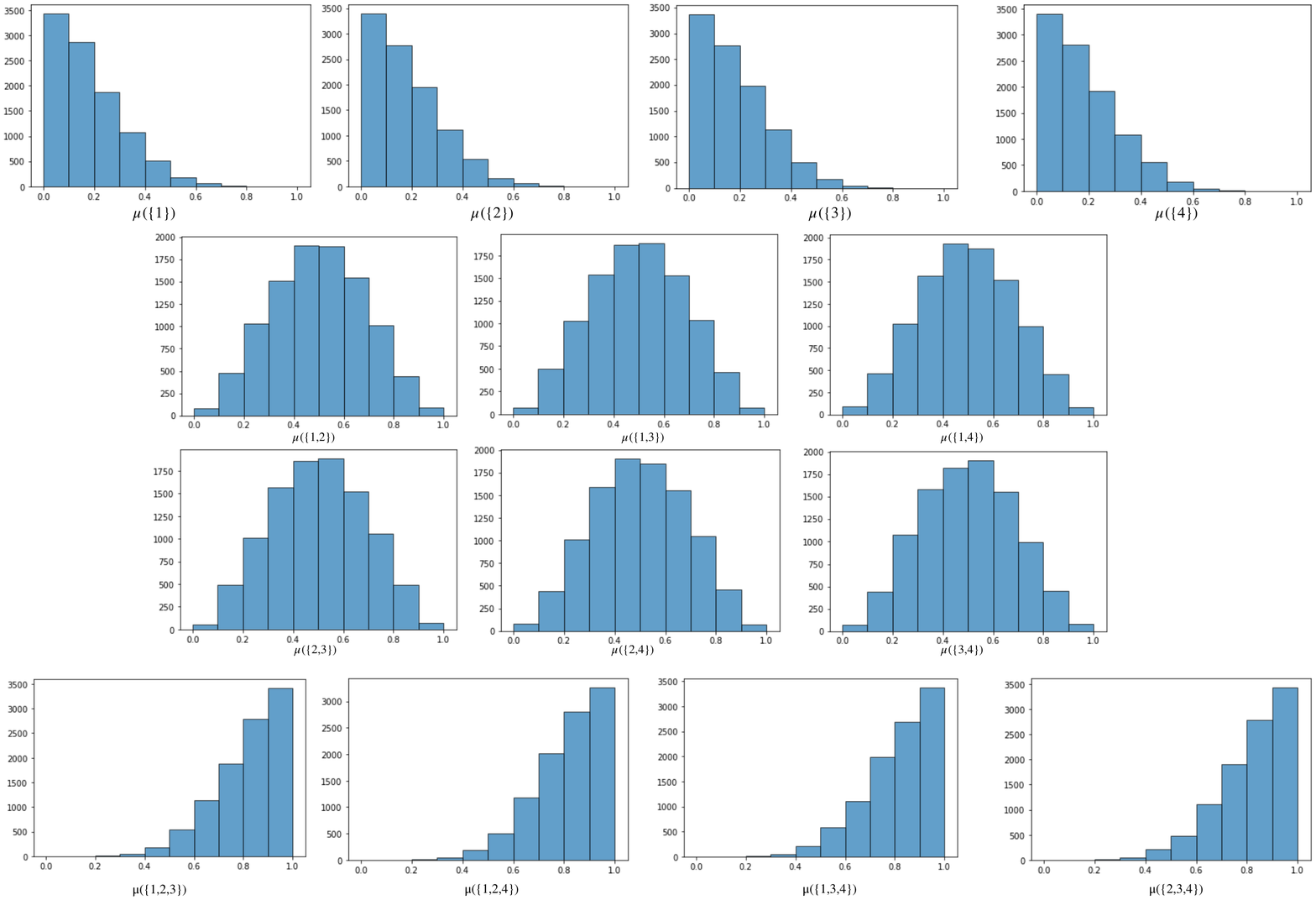}
\caption{Case $n=4$. Histograms of the values of $\mu(S)$,
$S\in 2^N\setminus\{N,\varnothing\}$, generated by the Markov chain generator (compare with Fig.~\ref{fig:eg}).}
\label{fig:mc}
\end{figure}

From Figure~\ref{fig:irn}, we notice that the distribution of $\mu$ generated by the
IRNG is much closer to the exact distribution than the one generated with the classical RNG (Fig.~\ref{fig:rn}), and Figure~\ref{fig:mc} shows the distribution of $\mu$ generated by the Markov Chain generator. 

\medskip
Next, we further compare their performance by calculating the Kullback-Leibler
divergence (also called Relative entropy) between the
distributions of $\mu(S)$ obtained by the exact generator and those obtained
by the considered generators, which could be used to estimate the similarity
of two distributions. ~\\ Recall the definition of Kullback-Leibler divergence:
$$\mathbb{D}_{KL}(p||q) = \sum_{x\in \mathcal{X}}p(x)\log\frac{p(x)}{q(x)}$$
with $p$ and $q$ two discrete probability distributions defined on the same probability space $\mathcal {X}$. 

In our experiments, we need to compare the distribution of $\mu(S)$ generated by the considered generators with the exact distribution of $\mu(S)$. We replace $q$ by the exact distribution of $\mu(S)$ and $p$ by the distribution of $\mu(S)$ obtained by one of these three generators and then compare their value. The smaller the value, the higher the similarity with the exact distribution (shown in Table~\ref{Table1}).

Table~\ref{Table2} shows the CPU time of different capacity
generators (we have used Python implementations of the algorithms described
above and have conducted the experiments on a 3.2 GHz PC with 16 GB of
RAM). For the execution time of IRNG, the time required to compute the
probabilities in Equation $(\ref{Eqq2})$ is not taken into account, as they
are computed once for all off line.

~\\
\begin{table}[H]
\centering
\begin{tabular}{|c|cccc|}
\hline
capacity generator & $\mu(\{1\})$ & $\mu(\{2\})$ & $\mu(\{3\})$ & $\mu(\{4\})$ \\
\hline
RNG & 0.4220 & 0.3651 & 0.3708 & 0.3947\\
IRNG & 0.0376 & 0.0392 & 0.0356 & 0.0367\\
Markov Chain & 0.0115 & 0.0109 & 0.0097 & 0.0073\\
\hline
\end{tabular}

~\\
~\\

\begin{tabular}{|c|cccccc|}
\hline
capacity generator & $\mu(\{1,2\})$ & $\mu(\{1,3\})$ &$\mu(\{1,4\})$ & $\mu(\{2,3\})$ & $\mu(\{2,4\})$ & $\mu(\{3,4\})$ \\
\hline
RNG & 0.6677 & 0.6322 & 0.7401 & 0.6522 & 0.6836 & 0.6375\\
IRNG & 0.0223 & 0.0187 & 0.0222 & 0.0253 & 0.0178 & 0.0191\\
Markov Chain & 0.0093 & 0.0090 & 0.0110 & 0.0108 & 0.0090 & 0.0102\\
\hline
\end{tabular}

~\\

~\\

\begin{tabular}{|c|cccc|}
\hline
capacity generator & $\mu(\{1,2,3\})$ & $\mu(\{1,2,4\})$ & $\mu(\{1,3,4\})$ & $\mu(\{2,3,4\})$ \\
\hline
RNG & 0.3985 & 0.3818 & 0.3691 & 0.3701\\
IRNG & 0.0296 & 0.0270 & 0.0270 & 0.0258\\
Markov Chain & 0.0089 & 0.0072 & 0.0081 & 0.0094\\
\hline
\end{tabular}
\caption{Kullback-Leibler divergence between the histograms produced by the
considered generators and those produced by the exact generator}
\label{Table1}
\end{table}
~\\

From Table~\ref{Table1}, we compute the sum of the Kullback-Leibler divergences
for $\mu(S)$ $(\forall S\in 2^N\setminus\{\varnothing,N\})$ for each
generator. We obtain that the value for RNG is $7.086$, for IRNG is $0.384$ and for Markov Chain is
$0.132$. As can be seen from these results, compared to the RNG, the
distribution of $\mu$ obtained from the IRNG is considerably improved and does
not differ significantly from the distribution obtained with the Markov chain generator.

\begin{table}[H]
\centering
\begin{tabular}{|c|c|c|}
\hline
Method & four elements' capacity & five elements' capacity\\
\hline
RNG& 0.425 & 1.130 \\
IRNG& 2.142 & 16.135 \\
Markov Chain Generator& 25.270& 243\\
\hline
\end{tabular}
\caption{Comparison of CPU time for generating 10000 capacities}
\label{Table2}
\end{table}

Unlike RNG, IRNG needs to compute $\Min_{\p}\mathcal{Rk}(S)$
and $\Max_{\p}\mathcal{Rk}(S)$ for each $S$. Therefore, IRNG is theoretically
more complex than RNG, and this difference is reflected in the computation
time. However, from Table~\ref{Table2}, this difference remains negligible in view of the time required by the Markov chain method, and it can be seen that IRNG is much faster than the Markov Chain Generator. This definitely shows
the advantage of IRNG, whose performance is dramatically better than that of
RNG, and not much different from that of the Markov Chain Generator.

%=======================================================================
%=======================================================================
\section{Capacity generators with preference information}
In this section, we would like to add some additional restrictions to the ECG
and IRNG, so that they could be applied in a multi-criteria decision problem
based on Choquet integral preference model.

\subsection{Statement of the problem}
Consider a discrete multiple criteria decision problem with $n$ criteria
where the decision is based on a Choquet integral, as {described} in Section \ref{Sback1}. The capacity is obtained from preference information provided by the DM, in terms of indifference statements $I$ and strict preference statements $P$ regarding some alternatives in a set of $m$ alternatives denoted
by $A$. 
% The decision maker (DM) could provide some preference information about alternatives, essentially strict preference and indifference denoted respectively by $\succ$ and $\sim$:
% \begin{itemize}
%	\item $I$ is the set of all ordered pairs $a$ and $b$ with indifference, formally $I := \{(a, b) \in A^2: a \sim b\}$;
%	\item  $P$ is the set of all ordered pairs $a$ and $b$ with strict preference, formally $P := \{(a, b) \in A^2: a \succ b\}$.
%\end{itemize}
%Then we use the Choquet integral to represent the DM's preference information
We denote these preference information constraints as the system $(S_{R})$:

$$(S_{R}) \quad \quad  \left\{ 
\begin{array}{l}
C_{\mu}(a) = C_{\mu}(b), \text{ with } (a,b) \in I\\
C_{\mu}(a) \geq C_{\mu}(b) + \epsilon, \text{ with } (a,b) \in P.
\end{array}\right.
$$
In the above system, we set the value of $\epsilon>0$ to be sufficiently small
to ensure that there is no equality implied by the system of inequalities.

The purpose of this section is to randomly generate a set of capacities
compatible with system $(S_R)$, in a uniform way. 

The simplest method for generating capacities compatible with $(S_{R})$ is acceptance and rejection, i.e., accepting only those
generated capacities which are compatible with the system $(S_{R})$. The ratio
of accepted capacities among all capacities is called the {\it acceptance
  rate}. Although this method is easily implementable, it will fail to succeed
in practice if the acceptance rate is too small.  In order to
study this issue more explicitly, we introduce the following notation:
\begin{itemize}
	\item We denote the volume of the polytope consisting of the capacities compatible with $(S_{R})$ as $V_{S_{R}}$, and the volume of capacities on $2^N$ as $V_{\mathcal{C}(N)}$.
	\item We denote by $r_n^a(S_R)$ the acceptance rate of generating an $n$ elements' capacity compatible with $(S_R)$, i.e.,  $r_n^a(S_R) = \frac{V_{S_{R}}}{V_{\mathcal{C}(N)}}$.
\end{itemize}

Since it does not exist a closed-form formula giving $V_{S_{R}}$, we cannot obtain a general conclusion about $r_n^a(S_R)$. Nevertheless, through the following example, we illustrate that the value of $r_n^a(S_R)$ could be quite low.

\begin{example}\label{main_example} Suppose there are three criteria $(n = 3)$
  and 6 alternatives $(m=6)$, with utility value as shown in the following table.
\begin{table}[H]
\centering
\renewcommand\arraystretch{1}
\caption{3 criteria alternatives}
\begin{tabularx}{0.8\linewidth}{|C|C|C|C|}
\Xhline{1.5\arrayrulewidth}
item  & criterion 1       & criterion 2   & criterion 3       \\
\Xhline{1.5\arrayrulewidth}
$a^1$  & 0.6 & 0.8 & 0.7 \\
$a^2$ & 0.7 & 0.1 & 0.8     \\
$a^3$ & 0.4 & 0.3  & 0.8    \\
$a^4$ & 0.4 & 0.9  & 0.7    \\
$a^5$ & 0.9 & 0.1  & 0.5    \\
$a^6$ & 0.9 & 0.4  & 0.3    \\
\Xhline{1.5\arrayrulewidth}
\end{tabularx}
\label{Talbe:main_example}
\end{table}
Suppose the DM has the following preference information: $P_1 =
\{(a^5,a^2),(a^2,a^3),(a^6,a^4)\}$, that is, $a^5 \succ a^2 \succ a^3$, and $a^6
\succ a^4$. The system of preference information becomes:
$$(S_{R_1}) \quad \quad \left\{
\begin{array}{l}
C_{\mu}(a^5) \geq C_{\mu}(a^2) + \epsilon \\
C_{\mu}(a^2) \geq C_{\mu}(a^3) + \epsilon \\
C_{\mu}(a^6) \geq C_{\mu}(a^4) + \epsilon .
\end{array}\right.
$$

The volume of the capacities compatible with $(S_{R_1})$ (computed by software
VINCI\footnote{https://www.multiprecision.org/vinci/home.html}) is 6.50e-03,
and the volume $V_{\mathcal{C}}(N)$  is 6.67e-02 for $n=3$. Hence, the
acceptance rate of the above example is $0.097$, which means that if we would
like to generate 10,000 capacities compatible with $S_{R_1}$, we need ultimately to
generate approximately $103,092$ general capacities, which shall take quite a
lot of time. 
\end{example}

The above example shows that it is necessary to revise the algorithms of IRNG
and ECG presented in section~\ref{sec:1} in order to increase the acceptance
rate by incorporating some constraints into the algorithms.  However, the problem
we encounter at this stage is that it would be too difficult to incorporate
constraints like $4\mu(\{1\}) \geq \mu(\{3\}) + 2\mu(\{1,3\}) + 10\epsilon$ (first
inequality of $(S_{R_1})$) in the algorithms, since the links among these three
variables $\mu(\{1\}),\mu(\{3\}),\mu(\{1,3\})$ are too complex.

\subsection{An approximation method}
Our idea is basically to handle constraints between at most two variables, and
limiting to very simple ones, in such a way that these rough constraints
determine a polytope including the ``true'' polytope determined by the
constraints $(S_R)$, and as small as possible. 

We propose to restrict to two types of constraints, namely {\bf constraint 1}
and {\bf constraint 2}.
$$
(S_{C}) \quad \quad \left\{
\begin{array}{l}
\mu(S) \leq \mu(S')+ \alpha, \text{ or }  \mu(S) \geq \mu(S') + \alpha , \text{ with }  S' , S\in 2^{N} \text{ and } \alpha \in \mathbb{R} \textbf{ (constraint 1)} \\
\mu(S) \leq c, \mu(S) = c \text{ or } \mu(S) \geq c, \text{ with $c$ constant and } S\in 2^{N} \textbf{(constraint 2)}
\end{array}\right.
$$

As it will become clear later, algorithm IRNG is able to handle both types of
constraints. However, the principle of ECG is to enumerate all linear
extensions, and then randomly select one of them. The value of $\mu(S)$ is not
involved in the process, therefore \textbf{constraint 2} would be impossible to add
directly in the algorithm, while \textbf{constraint 1} could only be added directly
when $\alpha = 0$.

We call generically $(S_C)$ a set of constraints of this
type. We explain below how to get a system $(S_C)$ from a system $(S_R)$, but we
can already draw some important considerations. Denote by $V_{S_C}$ the volume
of the polytope determined by the system $(S_C)$, supposing that the latter
polytope contains the polytope of capacities determined by $(S_R)$. As the two
polytopes differ in general, we still need to apply the acceptance and rejection
method to select only capacities compatible with $(S_R)$. Therefore, there is an
acceptance rate $r_n^a(S_C) = \frac{V_{S_R}}{V_{S_C}}$. If
$\frac{V_{S_R}}{V_{S_C}}$ is much larger than
$\frac{V_{S_{R}}}{V_{\mathcal{C}(N)}}$ (acceptance rate $r_n^a(S_R)$ of the naive
approach), then we have succeeded in reducing
considerably the computation time of the algorithm.

We distinguish two types of alternative.

\subsubsection{Case of binary alternatives.}

Binary alternatives are special alternatives which utility values are limited to 0 and 1, and are denoted by $a_B$ for $B\subseteq N$ -- see Section \ref{Sback1} and Example~\ref{e2} below.
Binary alternatives are quite useful in practice.
They represent efficient ways to elicit a capacity.
They are central in particular in the MACBETH approach \cite{bava12} and its extension to the Choquet integral \cite{maygralab09,maygralab10}.
These alternatives are cognitively simple to represent for a DM as they are only good or bad with respect to all criteria.

% An alternative is said to be {\it binary} when its utility values are limited to 0 and 1. Hence, each binary alternative is in correspondance with a subset $B\subseteq N$ of the set of criteria, corresponding to the criteria with utility value 1. We denote by $a_B$ the binary alternative corresponding to $B$ (see Example~\ref{e2}).

By (\ref{Eback2}), the value of the Choquet integral for alternative $a_B$ is simply
\[
C_\mu(a_B) = \mu(B).
\]
Therefore, the preferential information on binary alternatives induce only
constraints of the type $\mu(B)\leq \mu(B')$, i.e., constraints of type
{\bf constraint 1} with $\alpha=0$.

\begin{example}\label{e2}: Suppose there are three criteria and 5 binary
  alternatives shown in the following table. We use the Choquet integral to describe
  the preference of DM for these five alternatives. We obtain
  $C_{\mu}(a_{\{1,3\}}) = \mu(\{1,3\})$, $C_{\mu}(a_{\{1,2\}}) = \mu(\{1,2\})$,
  $C_{\mu}(a_{\{2,3\}}) = \mu(\{2,3\})$ and $C_{\mu}(a_{\{2\}}) =
  \mu(\{2\})$. Since $\mu(\{2\}) \leq \mu(\{2,3\})$,  $\mu(\{2\}) \leq
  \mu(\{1,2\})$, alternative $a_{\{2\}}$ is dominated by alternatives
  $a_{\{1,2\}}$ and $a_{\{2,3\}}$, it could be ranked directly as the least
  favourite alternative. 
  Suppose the DM prefers $a_{\{1,3\}}$ over
  $a_{\{1,2\}}$, then we have $\mu(\{1,3\}) \geq \mu(\{1,2\})$, which is of the type \textbf{constraint 1} with $\alpha = 0$. 

\begin{table}[H]
\centering \renewcommand\arraystretch{1}
\caption{3 criteria binary alternatives}
\label{table:binary}
\begin{tabularx}{0.8\linewidth}{|P{1cm}|C|C|C|}
\Xhline{1.5\arrayrulewidth}
item  & criteria 1       & criteria 2    & criteria 3  \\
\Xhline{1.5\arrayrulewidth}
$a_{\{1,3\}}$  & 1 & 0 & 1 \\
$a_{\{1,2\}}$ & 1 & 1 & 0  \\
$a_{\{2,3\}}$ & 0& 1 & 1   \\
$a_{\{2\}}$ & 0 & 1  & 0  \\
$a_{N}$ & 1 & 1  & 1  \\
\Xhline{1.5\arrayrulewidth}
\end{tabularx}
\end{table}
\end{example}

\subsubsection{Case of arbitrary alternatives}\label{Normalcase}

Now let us consider the case where the utility value is not limited to 0 or
1. Firstly, observe that
if two alternatives $a,b$ are indifferent, we have $C_{\mu}(a) = C_{\mu}(b)$. Through this equation, one variable $\mu(S)$ ($S \in 2^N$) could
be expressed in terms of the others in the equation and hence can be eliminated. Then,
without loss of generality, one may consider that we have only strict preference
information. In the same way, variable $\mu(N)=1$ is eliminated.

The basic idea to get a system $(S_C)$ of constraints which determines a
smallest possible polytope is to get the minimal and maximal values of {\it every}
non-eliminated variable $\mu(S)$ by linear programming, taking as constraints
those given by $(S_R)$ plus the monotonicity constraints, and similarly minimal
values and maximal values of differences $\mu(S)-\mu(S')$, for every possible
$S,S'$.  For example, the corresponding linear program to obtain the minimal
value of $\mu(S)$ for a fixed $S$ reads:

$$(LP_{S}^{min}) \quad \quad\left\{
\begin{array}{r}
 \Min \: \mu(S) \quad \quad \quad \quad \quad \quad \quad \quad \quad\\ 
\text{subject to } \quad  \quad
C_{\mu}(a) \geq C_{\mu}(b) + \epsilon \quad \text{ for } (a,b)\in P \\
\mu(S') \geq \mu(S''), \quad \quad  \quad  S' \supseteq S''  \\
\end{array}\right.
$$
with $\epsilon$ positive, $S', S'' \in 2^N \setminus \{\emptyset,
N\}$. Similarly, for obtaining other inequalities in the system $(S_C)$, we simply
replace the linear objective function with $\Max\:\mu(S)$, $\Min\:[\mu(S) -
  \mu(S')]$ and $ \Max\:[\mu(S) - \mu(S')]$ respectively, for all non-eliminated variables
$\mu(S),\mu(S')$. This yields at most  $(2^{n-1}-1)*(2^{n}-1)$ linear programs
to solve, whose values permit to get the system $(S_C)$. Denoting the minimum value of $\mu(S)$ by $v_{min}^{S}$, the
maximum value of $\mu(S)$ by $v_{max}^{S}$, the minimum value of $\mu(S) -
\mu(S')$ by $v_{min}^{S,S'}$ and the maximum value of $\mu(S) - \mu(S')$ by
$v_{max}^{S,S'}$, we obtain

$$(S_C) \quad \quad \left\{
\begin{array}{l}
v_{min}^{S,S'}\leq \mu(S) -  \mu(S')\leq v_{max}^{S,S'}, \quad \quad  \forall S,S'\in 2^N \setminus \{\emptyset , N\}\text{ and } S \neq S' \quad \textbf{ (constraint 1)}\\
v_{min}^S\leq \mu(S) \leq v_{max}^S,  \quad \quad \quad  \quad  \quad \quad \forall S\in 2^N \setminus \{\emptyset , N\} \quad  \quad   \quad  \quad  \quad \quad  \quad     \textbf{ (constraint 2)}\\
\end{array}\right.
$$

~\\
\textbf{Example \ref{main_example} continued:} \textit{We take $\epsilon = 0.001$ for the system $(S_{R_1})$, and then construct the family of linear programs to obtain the system $(S_{C_1})$. For example, for a given $S = \{1\}$, and  $S'= \{2\}$, we specify $\Min[ \mu(\{1\}) - \mu(\{2\})]$ as the linear objective function, and the linear program reads as follows :}

$$(LP_{(\{1\},\{2\})}^{min}) \quad \quad \left\{
\begin{array}{r}
\Min \: [\mu(\{1\}) - \mu(\{2\})]  \quad  \quad   \quad   \quad   \quad   \quad   \quad   \quad   \quad \\ 
\text{subject to }  \quad   \quad 
C_{\mu}(a^i) \geq C_{\mu}(a^j) + 0.001 \quad \text{ for } (a^i,a^j)\in P_1 \\
\mu(S) \geq \mu(S')   \quad   \quad   \quad \forall S \supseteq S' \\
\end{array}\right.
$$

\textit{To obtain the solution of the above linear program, we use the solver GUROBI,
and the result is $\Min [\mu(\{1\}) - \mu(\{2\})] = 0$, implying that $\mu\{1\}
- \mu\{2\}\geq 0$. Then we replace the objective function to obtain the remaining linear programs. After solving 42 linear programs, we obtain the following system of inequalities as the solution:}
$$ 
(S_{C_1}) \quad \left\{
\begin{array}{l}
\textcolor{blue}{ 0.2} \leq \mu\{1\} \leq 1; \quad\textcolor{black}{0 \leq \mu\{2\} \leq 1}; \quad\textcolor{black}{ 0 \leq \mu\{3\} \leq 1} \\

 \textcolor{blue}{0.2} \leq \mu\{1,2\} \leq 1; \quad  \textcolor{blue}{0.4}\leq \mu\{1,3\} \leq 1; \quad \textcolor{black}{0\leq \mu\{2,3\} \leq 1}\\
 
\textcolor{blue}{0} \leq \mu\{1\} - \mu\{2\} \leq 1;\quad \textcolor{blue}{ -0.25} \leq \mu\{1\}-\mu\{3\} \leq 1; \quad \textcolor{blue}{-0.8} \leq \mu\{1\}-\mu\{1,2\} \leq 0 \\ 

\textcolor{blue}{-0.5} \leq \mu\{1\}-\mu\{1,3\} \leq 0; \quad \textcolor{blue}{-0.4} \leq \mu\{1\} - \mu\{2,3\} \leq 1; \quad \textcolor{black}{ -1 \leq \mu\{2\} - \mu\{3\} \leq 1} \\

\textcolor{black}{-1 \leq \mu\{2\} - \mu\{1,2\} \leq 0};  \quad -1 \leq \mu\{2\} - \mu\{1,3\} \leq \textcolor{blue}{0}; \quad  \textcolor{black}{ -1 \leq \mu\{2\} - \mu\{2,3\} \leq 0} \\
 
-1 \leq \mu\{3\} - \mu\{1,2\} \leq \textcolor{blue}{0.25}; \quad  \textcolor{black}{-1 \leq \mu\{3\} - \mu\{1,3\} \leq 0}; \quad  \textcolor{black}{-1 \leq \mu\{3\} - \mu\{2,3\} \leq 0}  \\

\textcolor{blue}{-0.5} \leq\mu\{1,2\} - \mu\{1,3\} \leq \textcolor{blue}{0.6}; \textcolor{blue}{-0.33} \leq \mu\{1,2\} - \mu\{2,3\}\leq 1; \textcolor{blue}{-0.4} \leq \mu\{1,3\} - \mu\{2,3\} \leq 1 \\
\end{array}
\right.
$$

\textit{The numbers marked in blue are obtained from $(S_{R_1})$, and the numbers in black
are the monotonicity and normalization constraint of a capacity. Then we use the
software VINCI to compute the volume of $(S_{C_1})$, which is 1.75e-02, which
gives an acceptance rate $r^a_n(S_C)=0.371$. Compared with the original
acceptance rate $r^a_n(S_R)= 0.097$, it has been improved by a factor three.}

\subsection{ECG with preference information}\label{ECG}
Now we would like to incorporate the system $(S_C)$ to ECG, restricting to \textbf{constraint 1} with $\alpha = 0$. The principle of the ECG algorithm is to find the set of maximal elements at each step. If {there exist} element $S, S'\in 2^N$ with $S \not\subset S' $ and $\mu(S) \leq \mu(S')$, element $S$ should be ranked before element $S'$, hence element $S$ cannot be selected in the set of maximal elements till element $S'$ is removed from the poset. 

\begin{figure}[H]
\begin{minipage}{.5\linewidth}
\begin{tikzpicture}[scale = 1.5]
\node[gray,right] at (-1,2){$12$};
\draw[ thick, red](-0.75,1.9) -- (-0.95,2.1);
\draw[ thick, red](-0.95,1.9) -- (-0.75,2.1);
\node[right] at (0,2){$13$};
\node[right] at (1,2){$23$};
\node[right] at (0,3){$123$};
\draw[thick,=0.5](0,2) -- (0,3);
\draw[lightgray,=0.5](-1,2) -- (0,3);
\draw[thick,=0.5](1,2) -- (0,3);
\node[right] at (0,0.4){$\cdots$};
\node[right] at (1,0.4){$\cdots$};
\draw[lightgray,,=0.5](-1.3,1) -- (-1,2);
\draw[lightgray,,=0.5](-0.7,1) -- (-1,2);
\node[lightgray,left] at (-1.3,1){$13$};
\node[lightgray,left] at (-0.7,1){$23$};
\draw[lightgray,=0.5](-1.55,0) -- (-1.3,1);
\draw[lightgray,=0.5](-1.05,0) -- (-1.3,1);
\node[lightgray,left] at (-1.55,0){$1$};
\node[lightgray,left] at (-1.05,0){$23$};
\draw[lightgray,,=0.5](-0.95,0) -- (-0.7,1);
\draw[lightgray,,=0.5](-0.45,0) -- (-0.7,1);
\node[lightgray,right] at (-0.95,0){$2$};
\node[lightgray,right] at (-0.45,0){$13$};
\node[lightgray,right] at (-1.5,-0.25){$\cdots$};
\node[lightgray,right] at (-0.8,-0.25){$\cdots$};
\draw[thick,=0.5](-0.3,1) -- (0,2);
\draw[thick,=0.5](0.3,1) -- (0,2);
\node[right] at (-0.3,1){$12$};
\node[right] at (0.3,1){$23$};
\draw[lightgray,=0.5](0.7,1) -- (1,2);
\draw[thick,=0.5](1.3,1) -- (1,2);
\node[lightgray,right] at (0.7,1){$12$};
\draw[ thick, red](0.95,0.9) -- (0.75,1.1);
\draw[ thick, red](0.75,0.9) -- (0.95,1.1);
\node[right] at (1.3,1){$13$};
\end{tikzpicture} 
\end{minipage}
\begin{minipage}{0.5\linewidth}
\begin{tikzpicture}[scale = 1.5]
\node[gray,right] at (-1,2){$12$};
\draw[ thick, red](-0.75,1.9) -- (-0.95,2.1);
\draw[ thick, red](-0.95,1.9) -- (-0.75,2.1);
\node[right] at (0,2){$13$};
\node[right] at (1,2){$23$};
\node[right] at (0,3){$123$};
\draw[thick,=0.5](0,2) -- (0,3);
\draw[lightgray,=0.5](-1,2) -- (0,3);
\draw[thick,=0.5](1,2) -- (0,3);
\draw[lightgray,=0.5](-0.3,1) -- (0,2);
\draw[thick,=0.5](0.3,1) -- (0,2);
\node[gray,right] at (-0.3,1){$12$};
\draw[ thick, red](-0.05,0.9) -- (-0.25,1.1);
\draw[ thick, red](-0.25,0.9) -- (-0.05,1.1);
\node[right] at (0.3,1){$23$};
\draw[lightgray,=0.5](0.7,1) -- (1,2);
\draw[thick,=0.5](1.3,1) -- (1,2);
\node[gray,right] at (0.7,1){$12$};
\draw[ thick, red](0.95,0.9) -- (0.75,1.1);
\draw[ thick, red](0.75,0.9) -- (0.95,1.1);
\node[right] at (1.3,1){$13$};
\draw[lightgray,,=0.5](0.,0) -- (0.3,1);
\draw[thick,=0.5](0.55,0) -- (0.3,1);
\node[gray,right] at (0.,0){$12$};
\draw[ thick, red](0.25,-0.1) -- (0.05,0.1);
\draw[ thick, red](0.05,-0.1) -- (0.25,0.1);
\node[,right] at (0.55,0){$3$};
\draw[lightgray,,=0.5](1.,0) -- (1.3,1);
\draw[thick,=0.5](1.55,0) -- (1.3,1);
\node[gray,right] at (1.,0){$12$};
\draw[ thick, red](1.25,-0.1) -- (1.05,0.1);
\draw[ thick, red](1.05,-0.1) -- (1.25,0.1);
\node[,right] at (1.55,0){$3$};
\node[,right] at (0.2,-0.35){$\cdots$};
\node[,right] at (1.2,-0.35){$\cdots$};
\end{tikzpicture} 
\end{minipage}
\hfill
\caption{{Left constraint: $\mu(\{1,3\}) \geq \mu(\{1,2\}).$ Right constraint:  
$\mu(\{3\}) \geq \mu(\{1,2\})$ (which implies $\mu(\{1,3\}) \geq \mu(\{1,2\})$ and $\mu(\{2,3\}) \geq \mu(\{1,2\})$)}}
\label{ECG_a}
\end{figure}
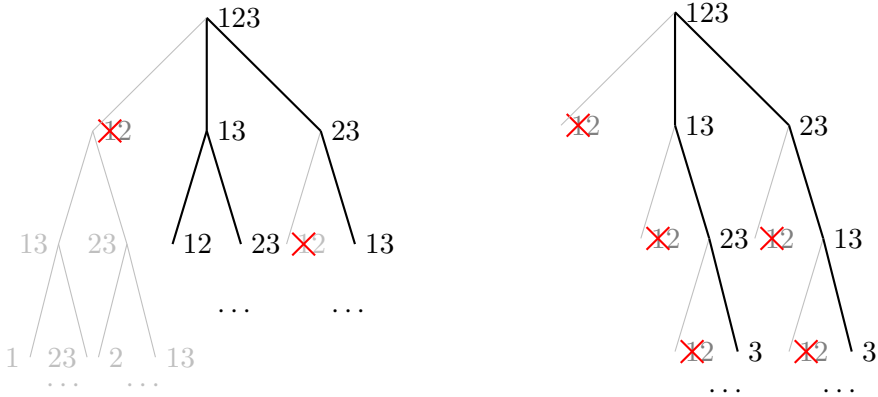

For example, if we add a constraint like $\mu(\{1,3\}) \geq \mu(\{1,2\})$ to the algorithm, at each stage when we find the set of maximal elements, $\{1,2\}$ cannot be in the set until $\{1,3\}$ has been {removed from the poset $(2^N, \subseteq$)} (shown in Figure \ref{ECG_a} left). If we add a constraint $\mu(\{3\}) \geq \mu(\{1,2\})$ into the algorithm, at each stage, element $\{1,2\}$ cannot be in the set until all the elements that contain $\{3\}$ have been {removed from the poset $(2^N, \subseteq$)} (shown in Figure \ref{ECG_a} right).

~\\
\textbf{Revised algorithm}

~\\ For the revised ECG algorithm, based on the original algorithm, we simply
modify the maximal set ($\Max(P)$) and add two arrays describing the DM's
preference information, called the {\it dominant} and the {\it dominated}
arrays. For a given preferential information $(a,b)\in P$, we put $a$ in the
dominant array and $b$ in the dominated array at the same position. For example,
if
we have $\mu(\{1\}) \leq \mu(\{2\})$ as the first restriction, then the first
element of the dominated array is {$\{1\}$}, and the first element of
the dominant array is {$\{2\}$}. Referring to {Example
  \ref{main_example}} (system $(S_{C_1})$), we could only add $\mu(\{1\}) \leq
\mu(\{2\})$ and $\mu(\{2\}) \leq \mu(\{1,3\})$ to ECG. Therefore, the dominant array
  would be $[\{2\},\{1,3\}]$ and the dominated array would be $[\{1\},\{2\}]$.
%$${\begin{tikzpicture}[] 
%	\draw(0,0) grid (2,1);
%	\draw (-1,0.5) node (2) {dominant : };
%	\draw (-0.2,0.5) node (4) {};
%	\node (from) {};	
%	\draw (2,2) node (f) {};
%	\draw (0.5,0.5) node (2) {$\{2\}$};
%	\draw (1.5,0.5) node (1) {$\{1,3\}$};
%  \end{tikzpicture} \qquad \qquad \begin{tikzpicture}[H]
%	\draw(0,0) grid (2,1);
%	\draw (-1,0.5) node (2) {dominated : };
%	\draw (-0.2,0.5) node (4) {};
%	\node (from) {};	
%	\draw (2,2) node (f) {};
%	\draw (0.5,0.5) node (2) {$\{1\}$};
%	\draw (1.5,0.5) node (1) {$\{2\}$};
%\end{tikzpicture}}$$
We denote by $S$ the set belonging to the \textit{dominated} array, and by $S'$
  the corresponding set in the \textit{dominant} array. This revised algorithm focuses on the fact that if $S \in \Max(P)$ and it exist $S'' \in \Max(P)$ with $S'' \supseteq S' $, then $S$ should be eliminated from $\Max(P)$. Compared to the original algorithm, the modifications are highlighted in blue.
\begin{quote}
{\bf All-linear-extension}$(P,AllLinearExtensions,count,dominated,dominant)$ \\
{\bf Input:} an array $P$ containing a poset of size $n$, an array  $AllLinearExtensions,count$, an array of dominated element and an array of dominant element\\
{\bf Output:} All constraint linear extensions of poset $P$\\
{\bf If} $|P| = 1$ {\bf then} \\
{\hspace*{1cm}} $\%$ When the bottom of a dendrogram is reached, add an empty linear extension to $AllLinearExtensions$.  \\
{\hspace*{1cm}} Append a zeros array of size $n$ to $AllLinearExtensions$ \\
{\hspace*{1cm}} $AllLinearExtensions[count-1][n-1] \leftarrow P[0]$  \\
{\hspace*{1cm}} $count \leftarrow count +1$ \\
{\bf end if}\\
\% Remove the dominated elements from $\Max(P)$.\\
\Old{{\bf For} $i$ in range of length dominated {\bf do}\\
{\hspace*{1cm}} {\bf For} $S$ in $\Max(P)$ {\bf do}\\
{\hspace*{2cm}} {\bf If} $S \supseteq dominant[i]$ and $dominated[i]$ in $\Max(P)$ : \\
{\hspace*{3cm}}  Remove dominated[i] from  $\Max(P)$\\
{\hspace*{3cm}}               break\\}
{\bf For} i in $ \Max(P)$ {\bf do}\\
{\hspace*{1cm}} Remove i from $P$\\
{\hspace*{1cm}} $\%$ recursion algorithm \\
{\hspace*{1cm}} {\bf All-linear-extension}($P,AllLinearExtensions,count,dominated,dominant$)\\
{\hspace*{1cm}} $AllLinearExtensions[count-1][\text{size of }P] \leftarrow$ i \\
{\hspace*{1cm}} Re-insert i to the end of poset $P$ \\
{\bf end for}\\
\end{quote}
~\\
\textbf{Experimental Results}

~\\
We compare the execution time of the revised and original ECG with acceptance
and rejection. For the experimental results, we add one by one the constraints
listed below, i.e., $\mu(\{1\}) \geq \mu(\{2\})$ the first time, then
$\mu(\{1,3\}) \geq \mu(\{4\})$ in addition the second time, etc. 
$$
\left\{
\begin{array}{l}
\mu(\{1\}) \geq \mu(\{2\}) \\
\mu(\{1,3\}) \geq \mu(\{4\}) \\   
\mu(\{2,3\}) \geq \mu(\{3,4\}) \\
\mu(\{1,2,3\}) \geq \mu(\{2,4\}) \\
\mu(\{1,2,4\}) \geq \mu(\{2,3,4\}) \\
\end{array}
\right.
$$
Table~\ref{table:rECG} gives the computation times required by the revised ECG
algorithm. They are to be compared with the computation time for the original
ECG with acceptance and rejection, which is always around 14s, not depending on
the number of constraints.

\begin{table}[h]
\centering
\begin{tabularx}{0.8\linewidth}{|C|C|C|C|C|C|}
\Xhline{1.5\arrayrulewidth}
number of restrictions  & 1 & 2 & 3 & 4 & 5 \\
\Xhline{1.5\arrayrulewidth}
Time  &  6.8s & 6.0s &  3.0s &  2.8s  & 1.4s \\
\Xhline{1.5\arrayrulewidth}
\end{tabularx}
\caption{Computational time for revised ECG when $n=4$}
\label{table:rECG}
\end{table}

The original ECG needs to generate all the linear extensions corresponding to poset $(2^N,\subseteq)$ and then filter them, whereas the revised algorithm only needs to generate the linear extensions that satisfy the constraints. Therefore, when we add more restrictions, the fewer the number of linear extensions, the less time the algorithm spends. 

\subsection{IRNG with preference information}\label{IRNG}

We present in this section the revised algorithm of IRNG when incorporating the
inequalities of the system $(S_{C})$, and illustrate it with Example \ref{main_example}. Throughout this section, we order the subsets of
$2^N\setminus\{\varnothing,N\}$ by the cardinal-lexicographic order, that is,
in increasing cardinality and using the lexicographic order for sets of same
cardinality (e.g., for $n=4$ we obtain 1, 2, 3, 4, 12, 13, 14, \ldots, 34, 123,
124, 134, 234, omitting braces and commas). We denote by $Ord(S)$ the rank of set $S$ in this order (e.g.,
$Ord(13) = 6$ when $n=4$). Hence we have always $Ord(\{1\})=1$ and
$Ord(N\setminus\{1\})=2^n-2$. We take the convention that in $v_{min}^{S,S'}$
and $v_{max}^{S,S'}$, always $Ord(S) < Ord(S')$.

Similar to ECG, the procedure of revised IRNG does not change, except that we
need to reconsider $\Max_{\p}\mu(S)$, $\Min_{\p} \mu(S)$,
$\Max_{\p}\mathcal{Rk}(S)$ and $\Min_{\p}\mathcal{Rk}(S)$ presented in Section~\ref{sec:thdi}. We introduce the following notation:
\begin{itemize}
\item Recall the value of already generated elements $S_1,\dots,S_p$ are $\mu(S_1) = a_1,\dots,\mu(S_p) = a_p$.
	\item We denote $\Max_{\p} \mu(S)$, $\Min_{\p} \mu(S)$ for revised IRNG by $\Max_{\p}^{r} \mu(S)$ $\Min_{\p}^{r} \mu(S)$.
	\item We denote the maximum and minimum values obtained from the inequalities of \textbf{constraint 1} in the system $(S_C)$ as $\Max_{\p}^{c1} \mu(S)$ and $\Min_{\p}^{c1} \mu(S)$, where:
\end{itemize}
{\[ \Min_{\p}^{c1}  \mu(S) =  \max_{ j\in \{1,\ldots,p\}}[(v_{min}^{S,S_j} + a_j)\mathbb{1}_{Ord(S) < Ord(S_j)}, (-v_{max}^{S_j,S} + a_j)\mathbb{1}_{Ord(S) > Ord(S_j)}] .
\]}
{\[ \Max_{\p}^{c1}  \mu(S) =  \min_{ j\in \{1,\ldots,p\}}[(v_{max}^{S,S_j} + a_j)\mathbb{1}_{Ord(S) < Ord(S_j)}, (-v_{min}^{S_j,S}+ a_j)\mathbb{1}_{Ord(S) > Ord(S_j)}] .
\]}
Then $\Min_{\p}^{r}\mu(S)$ and $\Max_{\p}^{r}\mu(S)$ becomes:
{\[ \Min_{\p}^{r}  \mu(S) =  \max(\Min_{\p} \mu(S), v_{min}^S,\Min_{\p}^{c1}  \mu(S)) .
\]}
{\[ \Max_{\p}^{r}  \mu(S) =  \min(\Max_{\p} \mu(S), v_{max}^S,\Max_{\p}^{c1}  \mu(S)) .
\]}
with $v_{min}^{S,S_j}$, $v_{max}^{S,S_j}$, $v_{min}^S$ and $v_{max}^S$ defined in section \ref{Normalcase}.

~\\
\textbf{Example \ref{main_example} continued:}\textit{ We incorporate the preference information into the algorithm in the form of the following two-dimensional array denoted as $MinMax$. Then we use the system $(S_{C_1})$ to show the complete process of computing $\Max_{\p}^{r}\mu(S)$ and $\Min_{\p}^{r}\mu(S)$.}
{\begin{table}[H]
\centering
\caption{Minimum and Maximum value of system (3)}
\begin{tabularx}{1.1\linewidth}{|C|C|C|C|C|C|C|C|C|C|}
\Xhline{2\arrayrulewidth}
  & $\mu(\{1\})$ & $\mu(\{2\})$ & $\mu(\{3\})$ & $\mu(\{1,2\})$ & $\mu(\{1,3\})$ & $\mu(\{2,3\})$ & $\mu(\{1\})-\mu(\{2\})$ & $\mu(\{1\})-\mu(\{3\})$& \dots \\
\Xhline{1.5\arrayrulewidth}
$\Min$ & 0.2  & 0 & 0  &  0.2  & 0.4 & 0 & 0 & -0.25 & \dots\\
$\Max$ &  1 & 1 & 1  &  1  & 1 & 1 & 1 & 1 & \dots\\
\Xhline{1.5\arrayrulewidth}
\end{tabularx}
\label{MinMax}
\end{table}} 
\textit{Let us suppose that we first randomly select element $\{2\}$ from $2^N\setminus \{N,\emptyset\}$ as the first one, and assign a value $a_1$ to it following the IRNG algorithm with $\Max^r\mu(\{2\}) = v_{max}^{\{2\}}$ and $\Min^r\mu(\{2\}) = v_{min}^{\{2\}}$. Then we randomly select the second element $\{1,3\}$, we need to find the value of $v_{min}^{\{1,3\}}$, $v_{max}^{\{1,3\}}$, $v_{min}^{\{2\},\{1,3\}}$ and  $v_{max}^{\{2\},\{1,3\}}$ from Table \ref{MinMax} which are 0.4, 1, $-1$ and 0, thus obtaining $\Min_{1}^r\mu(\{1,3\}) = \max(0.4,a_1)$, and $\Max_{1}^r\mu(\{1,3\}) = \min(1,1+a_1) = 1$. Then we assign a value $a_2$ to it following the IRNG algorithm and $a_2$ should be between $\max(0.4,a_1)$ and 1.}

{\begin{table}[H]
\centering
\begin{tabularx}{1.1\linewidth}{|C|C|C|C|}
\Xhline{1.5\arrayrulewidth}
   node & min & max & value \\
\Xhline{1.5\arrayrulewidth}
$\{2\}$ & 0  & 1 & $a_1$   \\
$\{1,3\}$ & $\max(0.4,a_1)$  & 1 & $a_2$ \\
$\cdots $ & $\cdots $ & $\cdots $ & $\cdots $\\
\Xhline{1.5\arrayrulewidth}
\end{tabularx}
\label{Example}
\end{table}} 
\textit{As this process continues, the subsequent element shall always be compared with each of the preceding ones to find $\Min_{\p}^r\mu(S)$ and $\Max_{\p}^r\mu(S)$.}

It is worth noting that if $v_{max}^{S, S'} \leq 0$ or $v_{min}^{S, S'} \geq 0$,
we obtain that the element $S$ should be ranked before element $S'$ or after
$S'$, respectively. We formalize the order relation between $\mu(S)$ and $\mu(S')$
by the quantity
\[
\mathcal{R}(S,S') =\begin{cases}
v_{max}^{S,S'}, & \text{ if } Ord(S) < Ord(S')\\
- v_{min}^{S',S}, & \text{ if } Ord(S) > Ord(S').
\end{cases}
\]
If  $\mathcal{R}(S,S') \leq 0$, then $\mu(S)\leq\mu(S')$. In such cases,
$\Min_{\p}\mathcal{Rk}(S)$ and $\Max_{\p}\mathcal{Rk}(S)$ should also be
recalculated. Taking Example \ref{Ex1} presented in Section~\ref{sec:thdi},
consider now $S = \{2,3\}$ with $N=\{1,2,3,4,5\}$, and suppose we have
$v_{max}^{\{1,3\},\{2,3\}} \leq 0$ from the system $(S_{C})$. Then the subset
  $\{1,3\}$ should be ranked before $\{2,3\}$. As we have $\mu(\{1,2\}) \leq
  \mu(\{1,3\})$ from the example, the subsets that are
  necessarily ranked before $S$ are the following: $\{1\},\{3\}, \{1,3\}, \{2\},\{1,2\}$. Then we apply (\ref{EqMinRk3}) to
  obtain the smallest possible ranking of $\{2,3\}$. To generalise
  $\overline{\mathcal{S}}_p(S)$, $\underline{\mathcal{S}}_p(S)$ with constraint (denoted as
  $\underline{\mathcal{S}}_p^c(S)$ and $\overline{\mathcal{S}}_p^c(S)$), we have

\[ \underline{\mathcal{S}}_p^c(S)=\{ S_j \ , \: j\in \{1,\ldots,p\}
\ \mbox{s.t. } \exists i \in \{1,\ldots,p\} \: , \{S_i \subseteq S \text{ or } \mathcal{R}(S_i,S) \leq 0\}  \mbox{ and } a_j \leq a_i\} \cup \{S\}
\]  is the set of already generated subsets that are necessarily ranked before $S$, and
\[ \overline{\mathcal{S}}_p^c(S)=\{ S_j \ , \: j\in \{1,\ldots,p\}
\ \mbox{s.t. } \exists i \in \{1,\ldots,p\} \:,  \{S_i \supseteq S \text{ or } \mathcal{R}(S,S_i) \leq 0\}  \mbox{ and } a_j \geq a_i  \} \cup \{S\}
\]is the set of already generated subsets that are necessarily ranked after $S$. 

Observe that the condition defining these collections are milder, therefore in
general $\underline{\mathcal{S}}_p^c(S)$ and $\overline{\mathcal{S}}_p^c(S)$ are
larger than the original ones without constraints.

In the end, we apply formulas (\ref{EqMinRk2}) and (\ref{EqMinRk3}) to obtain $\Max_{\p}\mathcal{Rk}(S)$ and $\Min_{\p}\mathcal{Rk}(S)$ with constraint denoted as $\Max^{r}_{\p}\mathcal{Rk}(S)$ and $\Min^{r}_{\p}\mathcal{Rk}(S)$. 

~\\
\textbf{Experimental result}

~\\
Since it takes more time to compute $\Min^{{r}}_{\p} \mu([S])$ and
$\Max^{{r}}_{\p} \mu([S])$ than in the original method, this revised algorithm
could be slower, as shown in Table \ref{Table_7}. Due to the difference between
the polytopes induced by $(S_{R})$ and $(S_{C})$, we still need to apply the
acceptance and rejection method after generating capacities by the revised
algorithm. For Example \ref{main_example} of Section \ref{Normalcase},
we generate 10,000 capacities compatible with the preference information
$(S_{R_1})$, and the computation times are shown in the following table.
{\begin{table}[H]
\centering
\caption{Computational time for generating 10,000 capacities in \textbf{Example \ref{main_example}}}
\begin{tabularx}{0.8\linewidth}{|C|C|C|}
\Xhline{1.5\arrayrulewidth}
  & revised IRNG & IRNG  \\
\Xhline{1.5\arrayrulewidth}
before acceptance and rejection  &  1.6s & 0.75s  \\
after acceptance and rejection  &  6.7s & 9.9s  \\
\Xhline{1.5\arrayrulewidth}
\end{tabularx}
\label{Table_7}
\end{table}}

From the table, we see that the revised IRNG algorithm takes about 2 times
longer to generate a capacity than the original one. Therefore, if the ratio of the two acceptance rates is more than two, the revised IRNG is faster than the original method. Recall that the original acceptance rate of the system $(S_{R_1})$ is around $10\%$ and the new acceptance rate of the system $(S_{R_1})$ from the system $(S_{C_1})$ is around $37\%$, hence the revised IRNG is faster than IRNG for Example \ref{main_example}.

For $n=4$, let us use the same example as for the revised ECG's experimental result, by gradually adding the following restrictions:
{$$
\left\{
\begin{array}{l}
\mu(\{1\}) \geq \mu(\{2\}) \\
\mu(\{1,3\}) \geq \mu(\{4\}) \\   
\mu(\{2,3\}) \geq \mu(\{3,4\}) \\
\mu(\{1,2,3\}) \geq \mu(\{2,4\}) \\
\mu(\{1,2,4\}) \geq \mu(\{2,3,4\}) \\
\end{array}
\right.
$$}
\\
The computation times for revised IRNG and original IRNG are shown in the following table.
\begin{table}[H]
\centering
\caption{Computational time for revised IRNG and IRNG when $n=4$}
\begin{tabularx}{0.8\linewidth}{|C|C|C|C|C|C|}
\Xhline{1.5\arrayrulewidth}
number of restrictions  & 1 & 2 & 3 & 4 & 5 \\
\Xhline{1.5\arrayrulewidth}
Revised IRNG  & 9.73s  & 11.66s  & 11.20s &  12.62s  & 12.07s \\
IRNG  & 7.86s  & 9.33s & 19.13s &  21.47s  & 45.00s \\
\Xhline{1.5\arrayrulewidth}
\end{tabularx}
\label{Table_8}
\end{table}
From the Table \ref{Table_8}, we notice that with the increase in the ratio of
the two acceptance  rates, revised IRNG is much faster than the original IRNG.

\section{Concluding remarks}
We have proposed an improved version of the random node generator of Havens and
Pinar, by investigating in a deeper way the probability distribution of the
coefficients $\mu(S)$. The results show that our algorithm yields distributions
much closer to the exact ones, compared to the original random node generator,
while keeping a very reasonable computation time, much smaller than the one
required by the Markov Chain method.

In a second part, we have incorporated simple constraints on the capacity
  coefficients into our algorithms. This permits to take into account some
  preferential information given by the Decision Maker. We have shown that we  
  can considerably improve the generation time compared to the naive approach of
  acceptance and rejection. 

Further studies will be devoted to the generation of special families of
capacities, like supermodular, $k$-additive capacities, etc.

\section{Data availability}
The datasets generated during and/or analysed during the current study are available from the
corresponding author on reasonable request.

\bibliographystyle{plain}
\bibliography{bib}
\end{document}